\documentclass[12pt]{article}
\usepackage{graphicx}
\usepackage[toc,page]{appendix}
\usepackage{amsmath}
\usepackage{mathrsfs}
\def\beq{\begin{equation}}
\def\eeq{\end{equation}}
\def\bea{\begin{eqnarray}}
\def\eea{\end{eqnarray}}
\def\nn{\nonumber}
\def\nl{\nonumber \\}

\def\scat{ \nu_{\tau}+ N \to \tau + X}
\def\scatmu{ \nu_{\mu}+ N \to \mu + X}

\def\nutau{ \nu_{\tau}}

\def\dec{\tau(p) \to \nu_{\tau}(p_3)+\pi^-(p_1)+ \pi^0(p_2)}

\def\BDtaunu{\bar{B} \to D^+ \tau^{-} \bar{\nu_\tau}} 
 
\def\BDlnu{\bar{B} \to D^+ \ell^{-} \bar{\nu_\ell}}
\def\BDstartaunu{\bar{B} \to D^{*+} \tau^{-} \bar{\nu_\tau}}

\def\BDstarlnu{\bar{B} \to D^{*} \ell^{-} \bar{\nu_\ell}}

\def\beq{\begin{equation}}
\def\eeq{\end{equation}}
\def\bea{\begin{eqnarray}}
\def\eea{\end{eqnarray}}
\def\bwt{\begin{widetext}}
\def\ewt{\end{widetext}}
\def\nn{\nonumber}
\def\roughly#1{\mathrel{\raise.3ex\hbox
{$#1$\kern-.75em\lower1ex\hbox{$\sim$}}}}

\def\bd{B^0}

\def\order{\lower 1.8ex \hbox{\LARGE\~{}}}

\def\BDtaunu{\bar{B} \to D^+ \tau^{-} {\bar\nu}_\tau}
\def\BDlnu{\bar{B} \to D^+ \ell^{-} {\bar\nu}_\ell}
\def\BDstartaunu{\bar{B} \to D^{*+} \tau^{-} {\bar\nu}_\tau}
\def\BDstarlnu{\bar{B} \to D^{*+} \ell^{-} {\bar\nu}_\ell}

\def \kstar{K^*}
\def\BKstarmumu{\bd \to \kstar \mu^+ \mu^-}

\def\ubar{\overline{u}}

\def\Qbar{\overline{Q}}

\def\psibar{\overline{\psi}}

\def\L{\mathcal{L}}

\begin{document}
 \unitlength = 1mm
\begin{flushright}
UMISS-HEP-2015-02 \\
[10mm]
\end{flushright}

\begin{center}
\bigskip {\Large  \bf Probing lepton non-universality in tau neutrino scattering}
\\[8mm]
Hongkai Liu $^{\dag}$
\footnote{E-mail:
\texttt{hliu2@go.olemiss.edu}}
, Ahmed Rashed $^{\dag\; \ddag\; \S}$  
\footnote{E-mail:
\texttt{amrashed@go.olemiss.edu}} 
 and Alakabha Datta $^{\dag}$ 
\footnote{E-mail:
\texttt{datta@phy.olemiss.edu}} 
\\[3mm]
\end{center}

\begin{center}
~~~{\it $^{\dag}$ Department  of Physics and Astronomy,}\\ 
~~~{ \it University of Mississippi,}\\
~~~{\it  Lewis Hall, University, Mississippi, 38677 USA}\\
\end{center}

\begin{center}
~~~{\it $^{\ddag}$ Department  of Physics, Faculty of Science,}\\ 
~~~{\it  Ain Shams University, Cairo, 11566, Egypt}\\
\end{center}

\begin{center}
~~~{\it $^{\S}$ Center for Fundamental Physics, Zewail City of Science}\\ 
~~~{\it  and Technology, 6 October City, Giza, Egypt}\\
\end{center}


\begin{center} 
\bigskip (\today) \vskip0.5cm {\Large Abstract\\} \vskip3truemm
\parbox[t]{\textwidth}  
{Recently hints of
 lepton flavor non-universality emerged in the BaBar and LHCb experiments. In this paper
 we propose tests of lepton universality in $\nu_{\tau}$ scattering. To parametrize the new physics we adopt an effective Lagrangian approach and consider the neutrino deep inelastic scattering processes $\scat$ and $\scatmu$
 where we assume the largest new physics effects are in the $\tau$ sector. We also consider an explicit leptoquark model in our calculations. In order to make comparison with the standard model and also in order to cancel out the uncertainties of the parton distribution functions, we consider the ratio of total and differential cross sections of tau-neutrino to muon-neutrino scattering.
 We find  new physics effects that can  possibly be observed at  the proposed Search for Hidden Particles (SHiP) experiment at CERN.
 }
\end{center}

\thispagestyle{empty} \newpage \setcounter{page}{1}
\baselineskip=14pt

\section{Introduction}

The flavor sector of standard model (SM) has many puzzles.  A key property of the SM gauge interactions is that they are lepton flavor universal. Evidence for violation of this property would be a clear sign of new physics (NP) beyond the SM. 
In the search for NP, the second and third generation quarks and leptons could 
be special because they are comparatively heavier and are 
expected to be relatively more sensitive to NP. 
As an example, in certain versions of the two Higgs doublet models (2HDM) the couplings of the new Higgs bosons are proportional to the masses and so NP effects are more pronounced for the heavier generations. Moreover, the constraints on new physics, especially involving the third generation leptons and quarks, are somewhat weaker allowing for larger new physics effects. 

Interestingly, there have been some reports of non-universality in the lepton sector from experiments. 
Recently, the BaBar
Collaboration with their full data sample has reported the
following measurements \cite{RDexpt1,RDexpt2}:
\begin{eqnarray}
\label{babarnew}
R(D) &\equiv& \frac{{\cal B}(\BDtaunu)}
{{\cal B}(\BDlnu)}=0.440 \pm 0.058 \pm 0.042 ~, \nn \\
R(D^*) &\equiv& \frac{{\cal B}(\BDstartaunu)}
{{\cal B}(\BDstarlnu)} = 0.332 \pm 0.024 \pm 0.018 ~,
\label{RDexpt}
\end{eqnarray}
where $\ell = e,\mu$. The SM predictions are $R(D) = 0.297 \pm 0.017$
and $R(D^*) = 0.252 \pm 0.003$ \cite{RDexpt1,RDtheory}, which deviate
from the BaBar measurements by 2$\sigma$ and 2.7$\sigma$,
respectively. (The BaBar Collaboration itself reported a 3.4$\sigma$
deviation from SM when the two measurements of Eq.~(\ref{babarnew})
are taken together.) This  measurement of lepton flavor non-universality,
referred to as the $R(D^{(*)})$
puzzles, may be providing a hint of the new physics (NP) believed to
exist beyond the SM.
There have been numerous analyses examining NP explanations of the
$R(D^{(*)})$ measurements \cite{dattaD,Bhattacharya:2014wla}.

In another measurement
the LHCb Collaboration recently measured the ratio of decay rates
for $B^+ \to K^+ \ell^+ \ell^-$ ($\ell = e,\mu$) in the dilepton
invariant mass-squared range 1 GeV$^2$ $\le q^2 \le 6$ GeV$^2$
\cite{RKexpt}. They found
\beq
R_K \equiv \frac{{\cal B}(B^+ \to K^+ \mu^+ \mu^-)}{{\cal B}(B^+ \to
  K^+ e^+ e^-)}
= 0.745^{+0.090}_{-0.074}~{\rm (stat)} \pm 0.036~{\rm (syst)} ~,
\eeq
is a $2.6\sigma$ difference from the SM prediction of $R_K = 1
\pm O(10^{-4})$ \cite{RKtheory}. 
In addition, we note that the three-body decay $\BKstarmumu$ by itself
offers a large number of observables in the kinematic and angular
distributions of the final-state particles, and it has been argued
that some of these distributions are less affected by hadronic
uncertainties \cite{BKmumuhadunc}. Interestingly, the measurement of
one of these observables shows a deviation from the SM prediction
\cite{Aaij:2013qta}. However, the situation is not clear whether this
anomaly is truly a first sign of new physics \cite{BKmumuNP}.

The tau neutrino, $\nutau$, was discovered by the DONuT experiment \cite{Kodama:2007aa} which measured the charged-current (CC) interaction cross section of the tau neutrino. The DONuT central-value results for the $\nu_{\tau}$ scattering cross section  show  deviation from the standard model predictions by about  40\% but with large experimental errors; thus, the measurements are consistent with the standard model. The third generation lepton has been explored relatively less than the other two generations and in particular there has not been much investigation of $\nutau$ properties. One of the predictions of the Standard Model (SM) is that gauge bosons couple to the three generations of leptons universally. A careful test of this prediction is very important and observation of non-universality in the interactions of the lepton families would be an important discovery. 

In previous publications we considered new physics in $\nu_{\tau}$ scattering for
quasi-exclusive, resonant and DIS scattering \cite{ourpapers}. In those papers we were more focused on the error in the extraction of neutrino mixing angles in presence of new physics. In this paper we  focus on observables that may be measured at a $\nu_{\tau}$ scattering experiment. There is a  
proposed Search for Hidden Particles (SHiP) experiment at CERN \cite{SHIP} which is expected to have a large sample of  tau neutrinos which could be used to probe new physics in $\nu_{\tau}$ scattering. In our previous work we did not include new physics tensor interactions which we consider in this work.
In this work we will be interested at neutrino energies where the DIS component of the scattering process is dominant.

We  start with an effective Hamiltonian description of new physics operators. We  fix the constraints on the couplings from charged current $\tau$ decays. We  consider the
decays $ \tau \to \pi \nu_{\tau}$ and  $ \tau \to  \pi \pi \nu_{\tau}$ which are well measured. We  finally consider an explicit leptoquark model where  both scalar and tensor interactions, with relations between the couplings of the two interactions, are present.

The paper is organized in the following manner. In sec. 2  we introduce the effective Lagrangian to parametrize the NP operators,  describe the formalism of the decay process and introduce the relevant observables. In sec.3 we present our results and in sec.4 we present our conclusions. We collect some of our equations in Appendices (A, B).

\section{Formalism}
\label{formalism}

In the presence of NP, the effective Hamiltonian for the scattering process
$\scat$ 
  can be written in the form \cite{ccLag},
\bea
{\cal{H}}_{eff} &=& \frac{4 G_F V_{ud}}{\sqrt{2}} \Big[ (1 + V_L)\,[\bar{u} \gamma_\mu P_L d] ~ [\bar{l} \gamma^\mu P_L \nu_l] \, +  V_R \, [\bar{u} \gamma^\mu P_R d] ~ [\bar{l} \gamma_\mu P_L \nu_l] \nl && \, + S_L \, [\bar{u} P_L d] \,[\bar{l}  P_L \nu_l] \, +  S_R \, ~[\bar{u} P_R d] \,~ [\bar{l}  P_L \nu_l]  \,  + T_L \, [\bar{u} \sigma^{\mu \nu} P_L d] \,~[\bar{l} \sigma_{\mu \nu} P_L \nu_l]\Big]\,,
\label{Heff}
\eea 
where  $G_F = 1.1663787(6) \times 10^{-5} GeV^{-2}$ is the Fermi coupling constant, $V_{ud}$ is the Cabibbo-Kobayashi-Maskawa (CKM) matrix element, $P_{L,R} = ( 1 \mp \gamma_5)/2$  are the projectors of negative/positive chiralities. We use $\sigma_{\mu \nu} = i[\gamma_\mu, \gamma_\nu]/2$ and
assume the neutrino to be always left chiral. To introduce non-universality the NP couplings are in general different for different lepton flavors. We assume the NP effect is mainly through the $\tau$ lepton.
The effective Hamiltonian involves the quarks of the first generations only. It is possible that the quarks of the other generations will also be affected by new physics. 
We will not assume any connection between new physics for the different generations of quarks.
The SM  effective Hamiltonian corresponds to $g_L = g_R = g_S = g_P = 0$.

The Hamiltonian in the presence  of only scalar and tensor operators can be written as,
\beq
{\cal {H}}_{\rm eff}=\frac{G_F V_{ud}}{\sqrt{2}}\left[ \bar{u}(A_S +B_S \gamma_5)d \; \bar{l} (1-\gamma_5)\nu_{l} +T_L \; \bar{u}\sigma^{\mu\nu}(1- \gamma_5)d \; \bar{l} \sigma_{\mu\nu}(1-\gamma_5)\nu_{l} \right],
\label{ST}
\eeq
where $A_S=S_R+S_L$ and $B_S=S_R-S_L$ with $S_L$ and $S_R$ are the left and right handed scalar couplings and $T_L$ is the tensor coupling.

We will first employ a model independent approach and treat the scalar and tensor couplings one at a time. Since,  in many realistic
models both the scalar and tensor couplings may be present,  we will consider an explicit leptoquark model where both the scalar and tensor couplings are present.

The Hamiltonian in the presence  of only $V \pm A$ operators  was considered in our previous work \cite{ourpapers}. There the effective Hamiltonian was written in terms of a $W'$ model, which could arise in extensions of the SM \cite{dattaKK}, as
\bea
{\cal{L}} &=& \frac{g}{\sqrt{2}}V_{ f^\prime f} \bar{f}^\prime \gamma^\mu( g^{f^\prime f}_L P_L +  g^{f^\prime f}_R P_R) f W^\prime_\mu + ~h.c.
\label{wprime}
\eea
 Integrating out the $W'$ leads to
\bea
{\cal{L}} &=& \frac{g^2}{2 M_{W}^2}V_{ f^\prime f} 
\left[\bar{f}^\prime \gamma^\mu\left(  \frac{M_W^2}{M_{W'}^2}g^{f^\prime f}_L P_L + \frac{M_W^2}{M_{W'}^2} g^{f^\prime f}_R P_R\right)  f \right] \left[g^{l, \nu_l}\bar{l} \gamma_\mu P_L \nu_l\right] + ~h.c.,\nl
{\cal{L}} &=&  \frac{4 G_FV_{ f^\prime f} }{\sqrt{2}}
\left[\bar{f}^\prime \gamma^\mu\left(  \frac{M_W^2}{M_{W'}^2}g^{f^\prime f}_L P_L + \frac{M_W^2}{M_{W'}^2} g^{f^\prime f}_R P_R\right)  f \right]\left [g_{\nu_l,l}\bar{l} \gamma_\mu P_L \nu_l \right ] + ~h.c. \nl
\label{intwprime}
\eea
Comparing Eq.~\ref{intwprime} with Eq.~\ref{Heff}  we have the following relations
\bea
V_L & = &\frac{M_W^2}{M_{W'}^2}g^{f^\prime f}_L g^{l, \nu_l}, \nl
V_R & = &\frac{M_W^2}{M_{W'}^2}g^{f^\prime f}_R g^{l, \nu_l}.
\label{LR}
\eea

\subsection{\fontsize{12}{10}\selectfont  Deep Inelastic Neutrino Nucleon Scattering}
In this section we discuss Deep Inelastic Neutrino Nucleon Scattering with the various types of interactions.

\subsubsection{\fontsize{12}{10}\selectfont  Scalar and Tensor Interactions}

In this section, we first present the total and differential cross sections for the deep inelastic scattering (DIS) process  
\bea
\scat,\;\;\;\scatmu,
\label{proc-DIS}
\eea
with scalar and tensor interactions. The total differential cross section is written in terms of contributions from the standard model, scalar and tensor operators and cross terms as follows
\beq
\frac{d\sigma_{\rm tot}}{dxdy}=\frac{d\sigma_{\rm SM}}{dxdy}+\frac{d\sigma_{\rm S}}{dxdy}+\frac{d\sigma_{\rm T}}{dxdy}+\frac{d\sigma_{\rm SM,ST}}{dxdy}+\frac{d\sigma_{\rm S,T}}{dxdy}.
\label{efflag}
\eeq
The differential cross section is given in terms of the cross section amplitude as follows
\beq
\frac{d\sigma_{\rm}}{dxdy}=\frac{1}{32\pi M E_\nu}\int \frac{d\xi}{\xi}f(\xi) |\bar{\mathcal{M}}(\xi)|^2 \delta(\xi-x).
\eeq
Here, $p_q^\mu = \xi p^\mu$ is the four-momentum of the scattered quark, $p^\mu$ is the target nucleon momentum, and $\xi$ is its momentum fraction. $f(\xi)$ is the the parton distribution function (PDF) inside a nucleon and $E_\nu$ is the incoming neutrino energy. In the deep inelastic scattering we calculate the differential cross section with respect to the scaling variables which are defined as follows
\bea
x &=& \frac{q^{2}}{2\,\nu},\nonumber\\ 
y &=& \frac{\nu}{M E_\nu},
\label{eq2}
\eea
where $x$ is the Bjorken variable and $y$ is the inelasticity with $q$ being the four-momentum transfer of the leptonic probe and
\beq
\nu=-p\cdot q=M(E_\nu - E_\ell).
\eeq
The physical regions for $x$ and $y$ are obtained by Albright and Jarlskog \cite{kretzer,albright}
\bea
\frac{m_{\ell}^{2}}{2M(E_{\nu}-m_{\ell})}\le x \le 1\label{aj1},
\eea
and
\bea
A-B \le y \le A+B,\label{aj2}
\eea
where
\bea
&&A=\frac{1}{2}\left(1-\frac{m_{\ell}^{2}}{2ME_{\nu}x}
-\frac{m_{\ell}^{2}}{2E_{\nu}^{2}}\right)
\bigg{/}\left(1+\frac{xM}{2E_{\nu}}\right),\label{aj3}\\
&&B=\frac{1}{2}\left[\left(1-\frac{m_{\ell}^{2}}{2ME_{\nu}x}\right)^{2}
-\frac{m_{\ell}^{2}}{E_{\nu}^{2}}\,\right]^{\frac{1}{2}}
\Bigg{/}\left(1+\frac{xM}{2E_{\nu}}\right). 
\label{aj4}
\eea

The terms in Eq.~\ref{efflag} are given as 
\bea
\frac{d\sigma_{\rm SM}}{dxdy}&=&\frac{G_F^2 M E_\nu}{\pi} \left(y(xy+\frac{m_\ell^2}{2ME_\nu})F_1+ (1-y-\frac{Mxy}{2E_\nu}-\frac{m_\ell^2}{4E_\nu^2})F_2\right.\nn\\
&&\left. + (xy(1-\frac{y}{2})-y\frac{m_\ell^2}{4ME_\nu})F_3 - \frac{m_\ell^2}{2ME_\nu} F_5 \right),\nonumber\\
\frac{d\sigma_{\rm S}}{dxdy}&=&\frac{G_F^2 M E_\nu}{4\pi}(A_S^2+B_S^2)y(xy+\frac{m_\ell^2}{2ME_\nu})F_1, \nonumber\\
\frac{d\sigma_{\rm T}}{dxdy}&=&\frac{8G_F^2 M E_\nu}{\pi}T_L^2 \left(y(xy+\frac{m_\ell^2}{2ME_\nu})F_1+ 2(1-y-\frac{Mxy}{4E_\nu}-\frac{m_\ell^2}{8E_\nu^2})F_2 -\frac{m_\ell^2}{ME_\nu}F_5 \right),\nonumber\\
\frac{d\sigma_{\rm SM,ST}}{dxdy}&=&0, \nonumber\\
\frac{d\sigma_{\rm S,T}}{dxdy}&=&\frac{2G_F^2 M E_\nu}{\pi} T_L (B_S-A_S) \left(xy(1-\frac{y}{2})-y\frac{m_\ell^2}{4ME_\nu}\right)F_3 . 
\eea
The functions $F_i$ are given as
\bea
F_1 &=& \sum_{q,\bar{q}}f_{q,\bar{q}} (\xi,Q^2) V^2_{q,q'},\nn\\
F_2 &=& 2\sum_{q,\bar{q}}\xi f_{q,\bar{q}} (\xi,Q^2) V^2_{q,q'},\nn\\
F_3 &=& 2\sum_{q}f_{q} (\xi,Q^2) V^2_{q,q'}-2\sum_{\bar{q}}f_{\bar{q}} (\xi,Q^2) V^2_{\bar{q},\bar{q}'},\nn\\
F_5 &=& 2\sum_{q,\bar{q}}f_{q,\bar{q}} (\xi,Q^2) V^2_{q,q'},
\eea
where $f_{q}$ and $f_{\bar{q}}$ are the parton distribution functions inside a nucleon, $V_{q,q'}$ is the CKM matrix element, and $Q^2=-q^2$.

One can write the differential cross sections above in terms of different variables $(t, \nu)$  using Eq.~\ref{eq2} and the transformation \cite{albright},
\beq
\frac{d\sigma}{dx dy}=2 M E_\nu \nu  \frac{d \sigma}{dq^2 d\nu}.
\label{trans}
\eeq
In the new variables, the differential cross sections can be written in the form 
\bea
\frac{d \sigma_{\rm SM}}{dq^2 d\nu} &=& \frac{G_F^2}{8\pi M E_\nu^2} \left(2(q^2 +m_\ell^2)W_1 + \left( 4E_\nu (E_\nu - \frac{\nu}{M})-(q^2+m_\ell^2) \right)W_2\right. \nonumber\\
&&\left. +\frac{1}{M^2}(2M E_\nu q^2-\nu (q^2+m_\ell^2))W_3 - \frac{2m^2_\ell E_\nu}{M} W_5  \right),\nonumber\\
\frac{d \sigma_{\rm S}}{dq^2 d\nu} &=& \frac{G_F^2}{16\pi M E_\nu^2} (A_S^2+B_S^2)(m_\ell^2 + q^2) W_1,\nonumber\\
\frac{d \sigma_{\rm T}}{dq^2 d\nu} &=& \frac{G_F^2}{\pi M E_\nu^2} T_L^2 \left( 2(m_\ell^2 + q^2) W_1 + (8 E_\nu^2 - (m_\ell^2 + q^2)-\frac{8E_\nu \nu}{M})W_2 - \frac{4 m_\ell^2 E_\nu}{M} W_5 \right),\nonumber\\ 
\frac{d \sigma_{\rm S,T}}{dq^2 d\nu} &=& \frac{G_F^2}{4\pi M^3 E_\nu^2} T_L (B_S-A_S) (2 E_\nu M q^2 -(m_\ell^2 + q^2)\nu)W_3 , 
\eea
where the (time-reversal invariant) structure functions are \cite{albright}
\bea
W_1(q^2,\nu)&=&\frac{F_1(x)}{M},\nonumber\\
W_2(q^2,\nu)&=&\frac{M F_2(x)}{\nu},\nonumber\\
W_3(q^2,\nu)&=&\frac{M F_3(x)}{\nu},\nonumber\\
W_5(q^2,\nu)&=&\frac{M F_5(x)}{\nu}.
\eea
We also define some Lorentz invariant variables in terms of the four-momenta of incoming
neutrino $(k)$, target nucleon $(p)$ and the produced charged lepton $(k')$ in the laboratory frame
\bea
&&Q^{2} = -q^{2}=-t,\\
&&W^{2} =(p+q)^{2}.
\eea
$Q^{2}$ is the magnitude of the momentum transfer and $W$ is the hadronic 
invariant mass. The physical regions of these variables are given by \cite{Hagiwara:2003di}
\bea
&&W_{\rm cut}\leq W \leq\sqrt{s}-m_{\ell},
\label{wregion}
\eea
in the DIS region with $W_{\rm cut}=1.4-1.6$ GeV, and
\bea
&&Q^{2}_{-}(W)\leq Q^{2}\leq Q^{2}_{+}(W),\label{qregion}
\eea
where $s=(k+p)^{2}$ and 
\bea
&&Q^{2}_{\pm}(W)=\frac{s-M^{2}}{2}(1\pm\bar{\beta})
-\frac{1}{2}\left[W^{2}+m^{2}_{\ell}
-\frac{M^{2}}{s}\left(W^{2}-m^{2}_{\ell}\right)\right],
\eea
with $\bar{\beta}=\lambda^{\frac{1}{2}}\left(1,m^{2}_{\ell}/s,W^{2}/s\right)$  
and $\lambda(a,b,c)=a^{2}+b^{2}+c^{2}-2(ab+bc+ca)$. In the lab frame, $s=M^2 + 2 M E_\nu$.

\subsubsection{\fontsize{12}{10}\selectfont  Explicit Leptoquark Model}

Here we will discuss an explicit  leptoquark model. 
Many extensions of the SM, motivated by a unified description of quarks and leptons, predict the existence of new scalar and vector bosons, called leptoquarks, which decay into a quark and a lepton. These particles carry non-zero baryon and lepton numbers, color and fractional electric charges. 
The most general dimension four  $SU(3)_c\times SU(2)_L\times U(1)_Y$ invariant Lagrangian of  leptoquarks satisfying baryon and lepton number conservation was considered in Ref~\cite{Buchmuller:1986zs}.
As the tensor operators in the effective Lagrangian get contributions only from scalar leptoquarks, we will focus only on scalar leptoquarks and consider the case where the leptoquark is a weak doublet or a weak singlet. The weak doublet leptoquark, $R_2$
has the quantum numbers   $(3,2,-7/6)$ under  $SU(3)_c\times SU(2)_L\times U(1)_Y$
while the singlet leptoquark $S_1$ has the quantum numbers $(\bar{3},1,-1/3)$.
 
The interaction Lagrangian that induces contributions to $\scat$ process is \cite{Queiroz:2014pra, laglq2}
\bea
\L_{2}^{\rm LQ} &=& \left( g_{2L}^{ij} \ubar_{iR}  L_{jL}  +g_{2R}^{ij} \Qbar_{jL} i\sigma_2 l_{iR}  \right) R_2, \nonumber\\
\L_{0}^{\rm LQ} & =& \left( g_{1L}^{ij},\Qbar_{iL}^c i\sigma_2 L_{jL} + g_{1R}^{ij},\ubar_{iR}^c \ell_{jR} \right)S_1, \
\label{leptoquark_lag}
\eea
where $Q_i$ and $L_j$ are the left-handed quark and lepton $SU(2)_L$ doublets respectively, while $u_{iR}$, $d_{iR}$ and $\ell_{jR}$ are the right-handed up, down quark and charged lepton $SU(2)_L$ singlets. Indices $i$ and $j$ denote the generations of quarks and leptons and $\psi^c = C\psibar^T=C\gamma^0\psi^*$ is a charge-conjugated fermion field.
The fermion fields are given in the gauge eigenstate basis
and one should make the transformation to the mass basis. Assuming the quark mixing matrices to be hierarchical, and considering only the leading contribution we can ignore the effect of mixing.

The couplings in Eq.~\ref{leptoquark_lag} can be constrained from $\tau$ decays. Because of the doublet nature of $R_2$ there will be additional term like $\bar{\tau} \tau \bar{q} q$ which do not contribute to $\tau$ decays. They will contribute to tau pair production but are much smaller than the SM production and hence do not add any new constraints.

After performing the Fierz transformations, one finds the general Wilson coefficients  at the leptoquark mass scale  contributing to the $\scat$ process:
\bea
   \label{LQ_WC}
      S_L &=& { 1 \over 2\sqrt2 G_F V_{ud} }  \left[ -{g_{1L}^{13}g_{1R}^{13*} \over 2M_{S_1}^2} - {g_{2L}^{13}g_{2R}^{13*} \over 2M_{R_2}^2} \right], \nonumber\\
      T_L &=& { 1 \over 2\sqrt2 G_F V_{ud} }  \left[ {g_{1L}^{13}g_{1R}^{13*} \over 8M_{S_1}^2} - {g_{2L}^{13}g_{2R}^{13*} \over 8M_{R_2}^2} \right]. \,
\eea
It is clear from Eq.~\ref{LQ_WC} that the combination from the weak singlet leptoquark and the weak  doublet can add constructively
or destructively to the Wilson's coefficients of the scalar and tensor operators in the effective Hamiltonian. 
In this section we will also consider the possibilities where both the scalar and the tensor operators are present and are of similar sizes. In the most general case both
the singlet and doublet leptoquarks are present and so both
 the scalar and tensor operators  appear in the effective Hamiltonian.
As there is limited experimental information, including both the singlet and the doublet leptoquarks will allow us more flexibility in fitting for the Wilson's coefficients
but this will come with the price of less precise predictions for the various observables.
We can, therefore, consider the simpler cases when only a singlet or a doublet leptoquark is present. In these cases, from Eq.~\ref{LQ_WC} the coefficients of scalar operators and the tensor operators are related. One can then consider the two  cases: 

Case. (a): In this case only the weak doublet scalar leptoquark  $R_2$
 is present.
In this case the Wilson's coefficients are
\bea
   \label{LQ_WC_a}
      S_L &=& { 1 \over 2\sqrt2 G_F V_{ud} }  \left[ - {g_{2L}^{13}g_{2R}^{13*} \over 2M_{R_2}^2} \right], \nonumber\\
      T_L &=& { 1 \over 2\sqrt2 G_F V_{ud} }  \left[ - {g_{2L}^{13}g_{2R}^{13*} \over 8M_{R_2}^2} \right]. \
\eea

Case. (b): In this case only the singlet leptoquark is present and the relevant Wilson's coefficients are
\bea
   \label{LQ_WC_b}
      S_L &=& { 1 \over 2\sqrt2 G_F V_{ud} }  \left[ -{g_{1L}^{13}g_{1R}^{13*} \over 2M_{S_1}^2}  \right], \nonumber\\
      T_L &=& { 1 \over 2\sqrt2 G_F V_{ud} }  \left[ {g_{1L}^{13} g_{1R}^{13*} \over 8M_{S_1}^2}  \right]. \,
\eea

In the  $(\bar 3, 1, -1/3)$ case, proton decay can occur in presence of an additional  leptoquark couplings to two quarks. The proton decay only constraints the leptoquark mass and the product of this additional coupling with the coupling considered here and so we can choose to turn the proton decay constraint as a constraint on the additional coupling
involving the two quarks.  Moreover, the relevant couplings for the considered processes involve the third generation and  so proton decay constraints do not apply in general to these couplings. Also note the proton cannot kinematically decay to a $\tau$.

The relations in Eqs.~(\ref{LQ_WC_a}, \ref{LQ_WC_b}) are valid at the leptoquark mass scale. We have to run them down to the $\tau$ mass scale using
the scale dependence of the scalar and tensor currents at leading logarithm approximation
\begin{eqnarray}
      S_L(m_\tau) &=& \left[ \alpha_s(m_t) \over \alpha_s(m_\tau) \right]^{\gamma_S \over 2\beta_0^{(5)}} \left[ \alpha_s(m_{\rm LQ}) \over \alpha_s(m_t) \right]^{\gamma_S \over 2\beta_0^{(6)}} S_L(m_{\rm LQ}), \nonumber \\ 
      T_L(m_\tau) &=& \left[ \alpha_s(m_t) \over \alpha_s(m_\tau) \right]^{\gamma_T \over 2\beta_0^{(5)}} \left[ \alpha_s(m_{\rm LQ}) \over \alpha_s(m_t) \right]^{\gamma_T \over 2\beta_0^{(6)}} T_L(m_{\rm LQ}) \,,
   \label{eq:QCDrunning}
\end{eqnarray}
where the anomalous dimensions of the scalar and tensor operators are $\gamma_S=-6C_F=-8$, $\gamma_T=2C_F=8/3$ respectively. Further, the beta function, $\beta_0^{(f)}=11-2n_f/3$ \cite{Dorsner:2013tla, Chetyrkin:1997dh, Gracey:2000am} and 
$n_f$ is the number of active quark flavors. 
One can use the equations above to run down the couplings from a chosen value of $m_{\rm LQ}$ to the tau mass, $m_\tau$.


In the presence of only one type of leptoquark, singlet or doublet state, one finds that the scalar $S_L$ and tensor $T_L$ Wilson coefficients are related to each other at the scale of leptoquark mass, $S_L(m_{\rm LQ})=\pm 4 T_L(m_{\rm LQ})$.

\subsubsection{\fontsize{12}{10}\selectfont  $V \pm A$ interactions}

The DIS differential cross section in the presence of V$\pm$A operators with respect to the variables $(x,y)$ is given in \cite{ourpapers}. Here we write it in terms of the momentum transfer, using Eq.~\ref{trans} as follows
\bea
\frac{d \sigma_{\rm SM+(V\pm A)}}{dq^2 d\nu} &=& \frac{G_F^2}{8 \pi M E_\nu^2}\left( \left(|a'|^2+|b'|^2 \right) (m_\ell^2 +q^2)W_1 \right. \nonumber\\
&& \left.  + \frac{1}{2M} \left(|a'|^2+|b'|^2\right) (4E_\nu^2 M -4 E_\nu \nu - M (m_\ell^2+q^2)) W_2 \right. \nonumber\\
&& \left. + \frac{1}{M^2} Re[a' b'^*] (2E_\nu M q^2- \nu (m_\ell^2+q^2)) W_3  - \frac{1}{M} \left(|a'|^2+|b'|^2 \right) m_\ell^2 E_\nu W_5 \right), \nonumber\\
\eea
where the definitions are 
\begin{eqnarray}
 a' &=& 1 + {\gamma}^\rho,\nonumber\\
 b' &=& 1 + {\gamma}^\kappa,\nonumber\\
 {\gamma}^\rho &=& V_L+V_R, \nonumber\\
 {\gamma}^\kappa &=& V_L-V_R.
\end{eqnarray}


%
%
%
%
%
%
%

\section{Constraints on NP couplings}

The scalar couplings $S_L$ and $S_R$ can be constrained by the tau decay channel $\tau^-(k_1) \rightarrow \nu_\tau (k_2)+\pi^-(q)$, while the tensor coupling $T_L$ can be constrained by the three-body decay channel  $\tau(p) \to \pi^-(p_1)+ \pi^0(p_2)+\nu_{\tau}(p_3)$. In this section we will discuss the constraints.

\subsection{$\tau^-(k_1) \rightarrow \nu_\tau (k_2)+\pi^-(q)$}

The hadronic current of the bound state $\pi$ can be parametrized as
\beq
\langle 0|\bar{d}\gamma^\mu (1-\gamma^5)u|\pi(q)\rangle = -i\sqrt{2}f_\pi q^\mu,
\label{SM-current}
\eeq
where $f_\pi=(92.4\pm 0.1\pm 0.3)$ MeV \cite{pdg} is the pion decay constant. The SM decay rate is
\beq
\Gamma_{\rm SM}^\pi=\frac{1}{8\pi}G_F^2 |V_{ud}|^2 f_\pi^2 m_\tau^3 \left(1-\frac{m_\pi^2}{m_\tau^2} \right)^2  \delta_{\tau/\pi}\,.
\eeq
Here $\delta_{\tau/\pi} =1.0016 \pm 0.0014$ \cite{rad} is  the radiative correction. Further, the SM  branching ratio can also be expressed as 
\cite{barish}
\begin{eqnarray}
Br^{SM}_{\tau^- \to \pi^- \nu_\tau}  &=&  0.607 Br(\tau^- \to \nu_\tau e^- \bar{\nu}_e) = 10.82 \pm 0.02 \% \,,
\end{eqnarray}
while the  measured $Br(\tau^- \to \pi^- \nu_\tau)_{exp} = (10.91 \pm 0.07)\%$ \cite{pdg}.
In the presence of a scalar state, the decay rate is
\beq
\Gamma_{\rm S}^\pi=\frac{1}{8\pi}G_F^2 |V_{ud}|^2 B_S^2 f_\pi^2 m_\pi^2 m_\tau \left(1-\frac{m_\pi^2}{m_\tau^2} \right)^2 ,
\eeq
where
\beq
\langle 0|\bar{d}(A_S-B_S \gamma^5)u|\pi(q)\rangle = i\sqrt{2}f_\pi m_\pi B_S.
\eeq
In order to obtain the scalar hadronic current above, we have multiplied the SM hadronic current (\ref{SM-current}) by the sum and difference of the quark momenta and used the equation of motion - see Appendix (B).
The total branching ratio can be written as follows
\beq
BR_{\rm tot}^\pi=BR_{\rm SM}^\pi\left(1+(r_{\rm S}^\pi)^2 \right), 
\eeq
where
\bea
(r_{\rm S}^\pi)^2 &=&\frac{BR_{\rm S}^\pi}{BR_{\rm SM}^\pi},
\eea
with
\bea
r_{\rm S}^\pi =\frac{B_S m_\pi}{m_\tau}.
\eea
Note, the interference term of the SM and the scalar NP term vanishes.

\begin{figure}[h!]
\centering
\includegraphics[width=7cm]{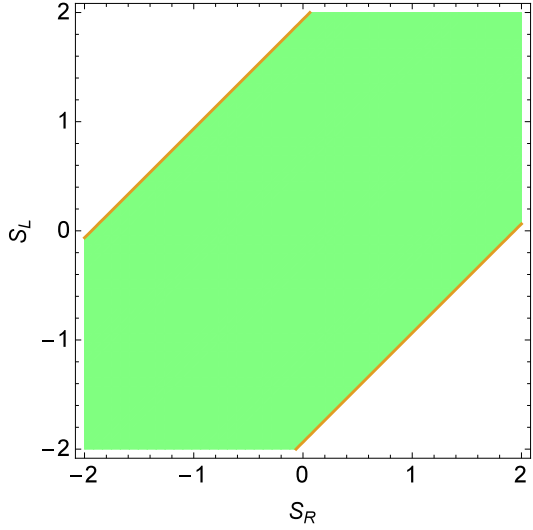}~~~
\caption{The constraints on the scalar couplings $S_{L,R}$. The colored region is allowed. The constraint is from $\tau^- \rightarrow \pi^- \nu_\tau$. We treat $S_L$ and $S_R$ as real couplings.}
\label{Const1}
\end{figure}

\begin{figure}[h!]
\centering
\includegraphics[width=7cm]{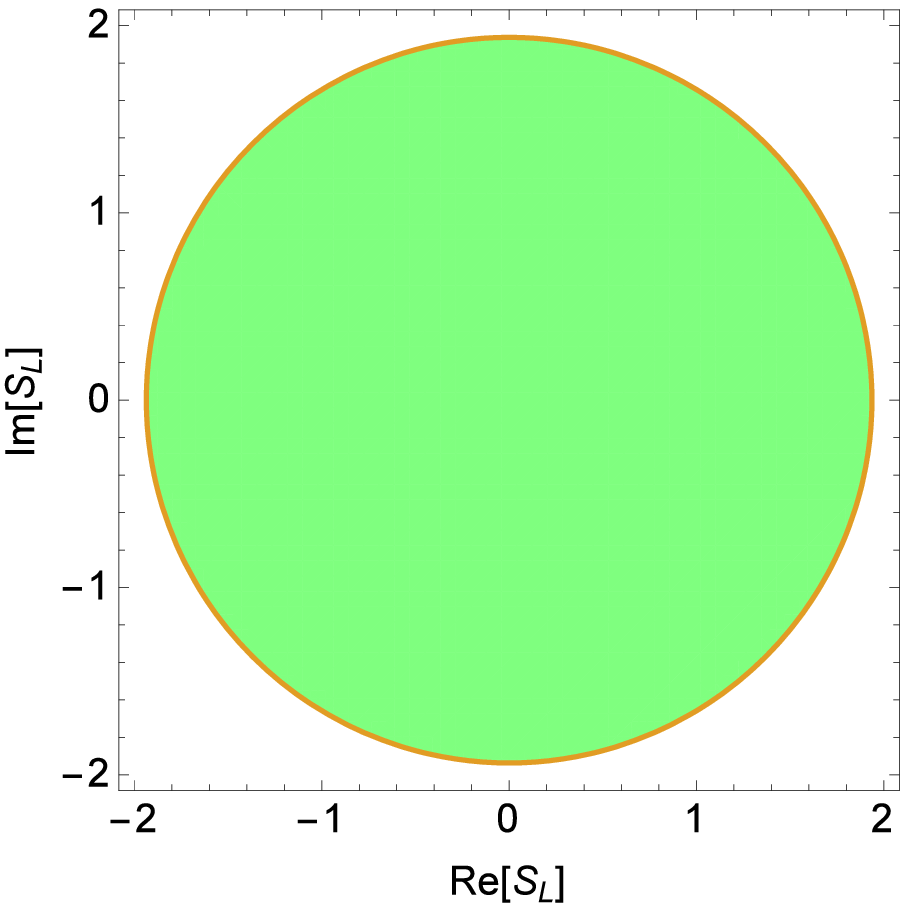}~~~
\includegraphics[width=7cm]{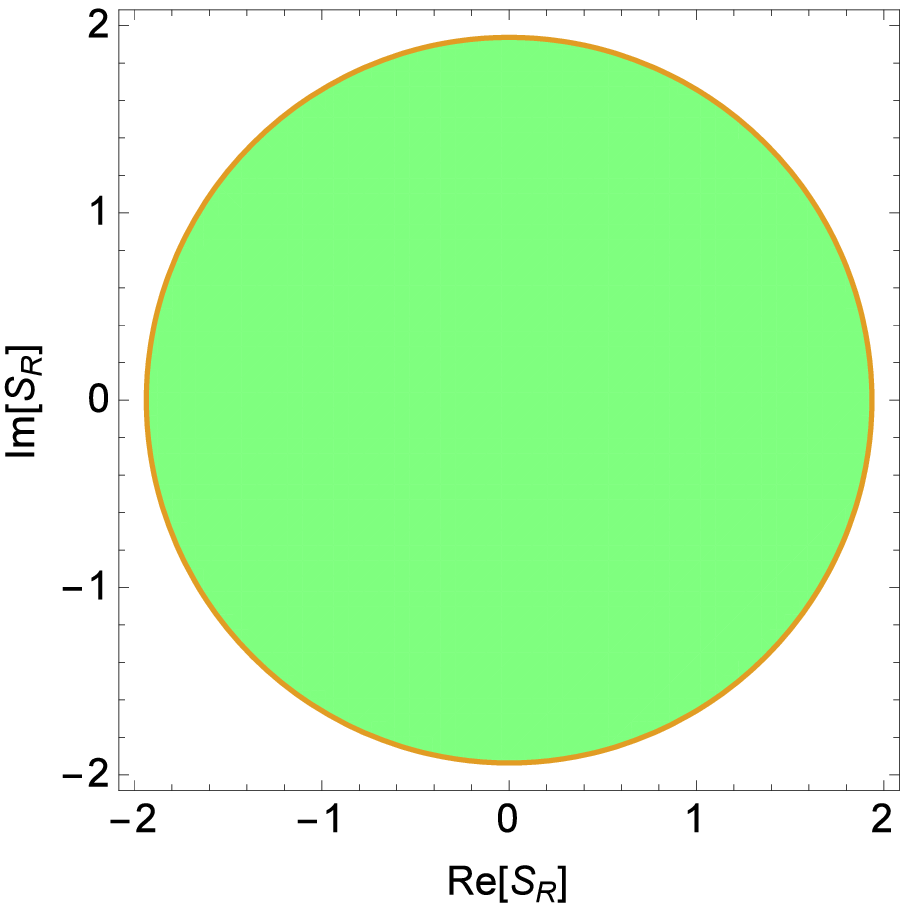}~~~
\caption{The constraints on the scalar couplings $S_{L,R}$. The colored region is allowed. The constraint is from $\tau^- \rightarrow \pi^- \nu_\tau$. Left panel: we take $S_R = 0$ and treat $S_L$ as a complex coupling.  Right panel: we take $S_L = 0$ and treat $S_R$ as a complex coupling.}
\label{Const11}
\end{figure}

The allowed region of the couplings are given in the contour plot Fig.~\ref{Const1} for the measured $\tau^- \to \pi^- \nu_\tau$ within the $2\sigma$ level. We consider theoretical uncertainty within the $1\sigma$ level. 
First, we assume both couplings, $S_{L,R}$, are present and take the couplings to be real. Next we assume the couplings are complex and take one coupling at a time, as shown in Fig.~\ref{Const11}. 

\subsection{$\tau(p) \to \pi^-(p_1)+ \pi^0(p_2)+\nu_{\tau}(p_3)$}

Here we consider two-pion decays of $\tau$. The process is
\begin{equation}
\dec.
\end{equation}
The SM and NP amplitudes are
\begin{equation}
M_{SM}=\frac{-iG_F V_{ud}}{\sqrt{2}}\langle \pi^-\pi^0|\bar{d}\gamma^\mu (1-\gamma^5)u|0 \rangle \bar{u}_{\nu_{\tau}}\gamma_{\mu}(1-\gamma^5)u_{\tau},
\label{MSM}
\end{equation}
\begin{equation}
 M_{T}=\frac{-iG_F V_{ud}}{\sqrt{2}}T_L\langle \pi^-\pi^0|\bar{d}\sigma^{\mu\nu} (1-\gamma^5)u|0 \rangle \bar{u}_{\nu_{\tau}}\sigma_{\mu\nu}(1-\gamma^5)u_{\tau}.
\label{MLQT}
\end{equation}
We can parametrize the relevant form factors as,
\begin{equation}
 \langle \pi^-\pi^0|\bar{d}\gamma^\mu (1-\gamma^5)u|0 \rangle=\sqrt{2}F(Q^2)k^{\mu},
 \label{current}
\end{equation}\\
\begin{equation}
\langle \pi^-\pi^0|\bar{d}\sigma^{\mu\nu} (1-\gamma^5)u|0 \rangle=\sqrt{2}F_T(Q^2)(k^{\mu}q^{\nu}-q^{\mu}k^{\nu}),
\label{current2}
\end{equation}\\
where $k=p_1-p_2$ and  $q=p_1+p_2$.
The form factor $F(Q^2)$, along with its error, is given in \cite{tau-decay, Bernicha:1995rh}. 
In our analysis, errors of the form factor parameters have been considered and included in  the constraint plots.
The origin of $\sqrt{2}$ comes from the wavefunction of  $\pi^0=\frac{1}{\sqrt{2}}(u\bar{u}+d\bar{d}).$ Considering the isospin symmetry, $u\bar{u}=d\bar{d}=\phi$, so $\pi^0=\sqrt{2}\phi.$ 
Using the equations of motion 
and by multiplying the SM hadronic current (\ref{current}) by $k^\nu$ and $q^\nu$, see Appendix (B),  we have
\begin{equation}
F_T=\frac{-i F}{\sqrt{q^2}}.
\end{equation}

One can find the details of the decay rate calculations in Appendix (A). We find that $\Gamma_{SM}=5.5\times 10^{-13}$ GeV. The total decay rate of $\tau$ is $\Gamma_{tot}=2.27\times 10^{-12}$ GeV, so that $BR(\tau^-\rightarrow\nu_{\tau}+\pi^-+\pi^0)$ is 24.23\% in our calculations which is close to the experimental result $(25.52\pm 0.09)\%$ \cite{pdg2pi}. Using the CVC hypothesis, it is predicted that $BR(\tau^- \rightarrow \pi^- \pi^0 \nu_\tau) = (24.75\pm 0.38) \%$ \cite{Bernicha:1995rh}.

\begin{figure}[h!]
\centering
\includegraphics[width=8.6cm]{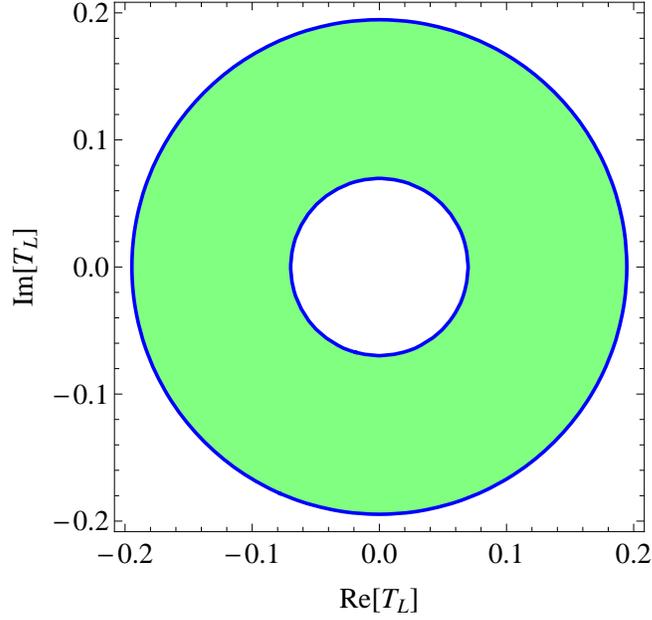}~~~
\caption{The allowed region for the real and imaginary components of the complex leptoquark coupling $T_L$. The constraint on $T_L$ is from $\tau^- \rightarrow \pi^-  \pi^0  \nu_\tau $.}
\label{SLmtauTLmtaurun2}
\end{figure}

From the constraint $\tau^- \rightarrow \pi^- \pi^0 \nu_\tau$, we find that $0.07 <|T_L| < 0.2$ within the experimental $2\sigma$ level while the theoretical uncertainty is considered within $1\sigma$. If we take the tensor coupling to be complex, the contour plot in Fig.~\ref{SLmtauTLmtaurun2} shows the allowed region of the real and imaginary components of the coupling for the measured $\tau \to \pi^- \pi^0 \nu_{\tau}$ within the experimental $2\sigma$ level and the theoretical uncertainty within $1\sigma$. The SM expectation for the branching ratio is not allowed within the experimental range at the $2\sigma$ level but it is allowed at higher standard deviation level. 

\begin{figure}[h!]
\centering
\includegraphics[width=6.5cm]{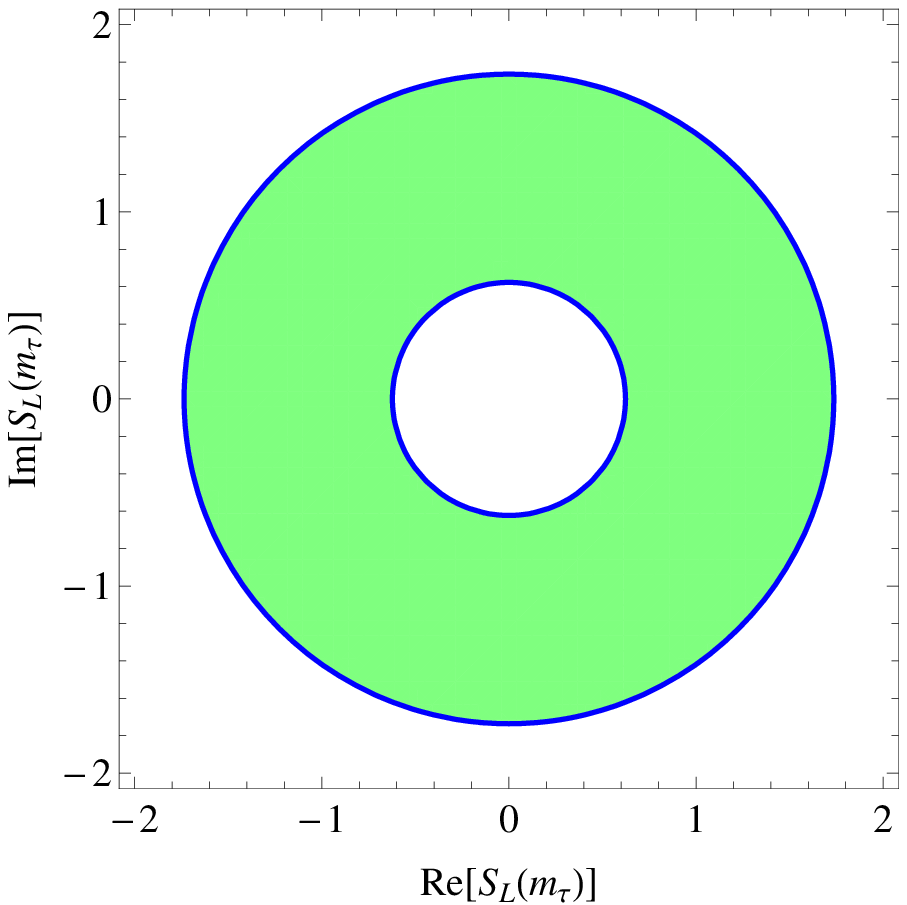}~~~
\includegraphics[width=6.8cm]{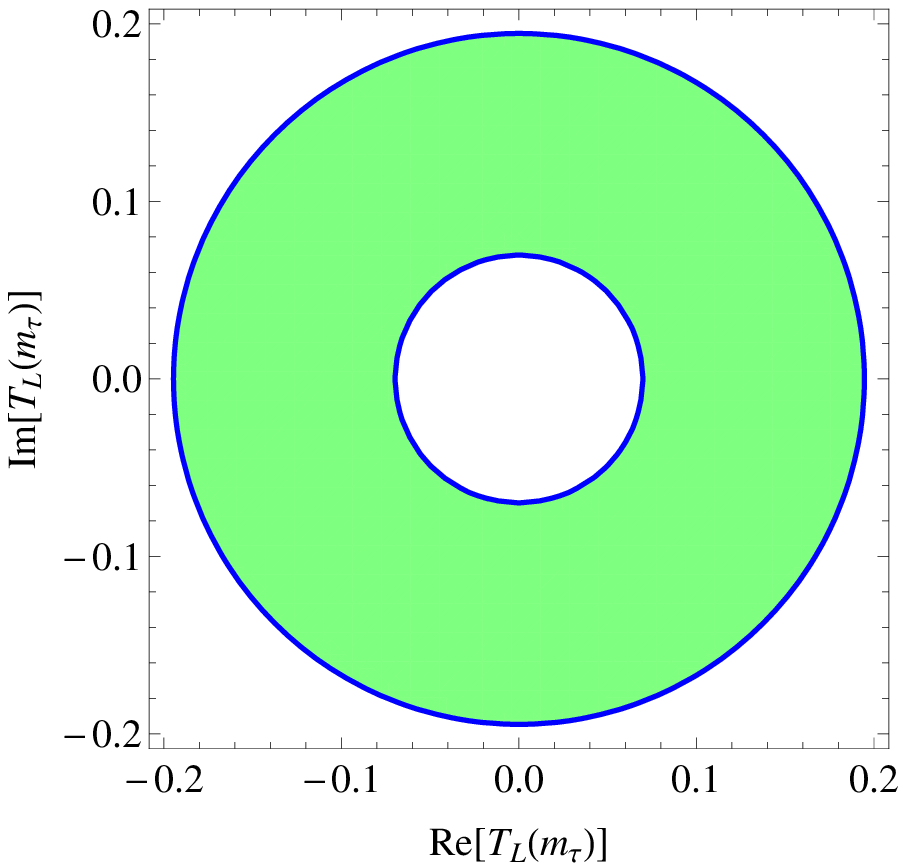}~~~
\caption{The allowed regions for the real and imaginary components of the leptoquark running couplings $S_L (m_\tau)$ and $T_L (m_\tau)$ with $S_L (m_{LQ})=\pm 4 \; T_L (m_{LQ})$ at $m_{\rm LQ}=1000$ GeV. The constraint on $S_L (m_\tau)$ is from $\tau^- \rightarrow \pi^- \nu_\tau$ and $T_L (m_\tau)$ is from $\tau^- \rightarrow \pi^- \pi^0 \nu_\tau $.}
\label{SLmtauTLmtaurun}
\end{figure}

In the explicit  leptoquark models, $S_L(m_{\rm LQ})=\pm 4 T_L(m_{\rm LQ})$, one can obtain the constraint on $S_L$ and $T_L$ from $\tau^- \rightarrow \nu_\tau \pi^-$ and  $\tau^- \rightarrow \pi^-  \pi^0  \nu_\tau$ at the same time. It is found that the limits of $0.07 <|T_L| < 0.2$ and $0.62 <|S_L| < 1.73$ are obtained within the experimental $2\sigma$ level with the theoretical uncertainty within $1\sigma$. The allowed regions of the real and imaginary components are shown in the contour plot in Fig.~\ref{SLmtauTLmtaurun}.

The $\pi \pi$ state is produced dominantly though an intermediate vector resonance and is in a P wave. Therefore, the scalar terms with the couplings $S_L$ and $S_R$ do not contribute to the decay process $\tau^- \rightarrow \nu_\tau \pi^-$ as the scalar hadronic current vanishes because of parity. Isospin symmetry also results in $S_{L,R}$ not contributing within decay.  


\begin{figure}[h!]
\centering
\includegraphics[width=6.5cm]{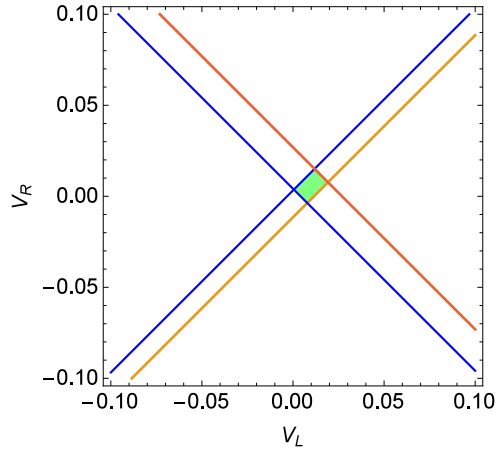}~~~
\caption{The allowed region (green area) for the left- and right-handed couplings $V_L$ and $V_R$. The constraints are from $\tau^- \rightarrow \pi^- \nu_\tau$ and  $\tau^- \rightarrow \pi^- \pi^0 \nu_\tau $.}
\label{ConstVpmA}
\end{figure}

\begin{figure}[h!]
\centering
\includegraphics[width=7cm]{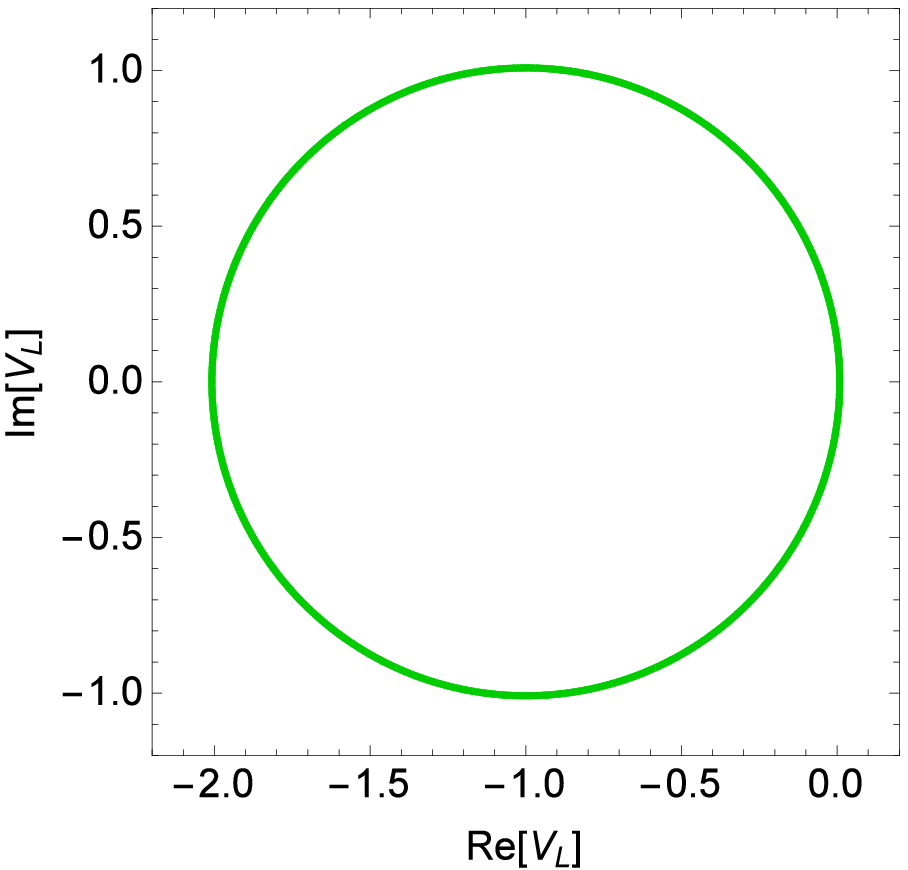}~~~
\includegraphics[width=7cm]{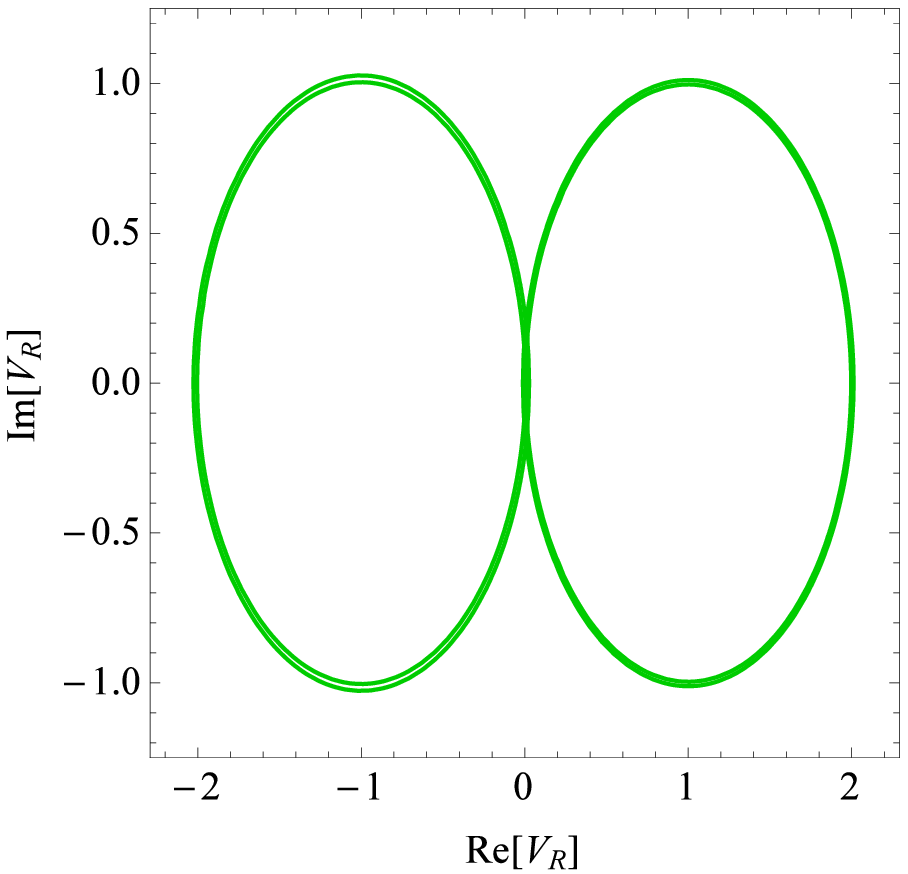}~~~
\caption{The allowed regions for the left- and right-handed complex couplings $V_L$ and $V_R$. The constraint are from $\tau^- \rightarrow \pi^- \nu_\tau$ and  $\tau^- \rightarrow \pi^- \pi^0 \nu_\tau $. Left panel: we take $V_R = 0$ and treat $V_L$ as a complex coupling. The allowed region is the contour ring.  Right panel: we take $V_L = 0$ and treat $V_R$ as a complex coupling. The allowed region is the overlap area between the two contour rings.}
\label{Const111}
\end{figure}

In the $V\pm A$ case, the couplings can be constrained by both $\tau^- \rightarrow \nu_\tau +\pi^-$ and $\tau \to \pi^-+ \pi^0+\nu_{\tau}$ decays. Considering the first process the branching ratio is given as 
\begin{eqnarray}
Br^{\pi}_{V\pm A} &=&  Br^{\pi}_{\rm SM}  (1+ r^\pi_{V\pm A})^2\,,
\end{eqnarray}
where the  $V\pm A$  contribution is  
\begin{eqnarray}
r^\pi_{V \pm A}  &=&  V_L - V_R\,.
\label{rpi}
\end{eqnarray}
{}From the second process, the branching ratio is given as
\begin{eqnarray}
Br^{\pi\pi}_{V\pm A} &=&  Br^{\pi\pi}_{\rm SM}  (1+ r^{\pi\pi}_{V\pm A})^2\,,
\end{eqnarray}
where the  $V\pm A$  contribution is  
\begin{eqnarray}
r^{\pi\pi}_{V \pm A}  &=&  V_L + V_R\,.
\label{rpi2}
\end{eqnarray}
If we take the couplings to be real, the contour plot in Fig.~\ref{ConstVpmA} shows the allowed region. The allowed regions for the real and imaginary parts, if the couplings are taken to be complex, are shown in the contour plot in Fig.~\ref{Const111}.

Note, even though we consider complex couplings in the constraint equations in this section, we take the couplings to be real for the scattering calculations.

\section{Numerical analysis}

In this section the sensitivity of the neutrino cross-section scattering to the scalar and tensor interactions, explicit Leptoquark model, and $V\pm A$ interactions is discussed. We study the ratio of the total cross section, $d \sigma/dxdy$, and $d\sigma/dt$ for the tau-neutrino to the muon-neutrino scattering. We also show the results of the total cross section, $d \sigma/dxdy$, and $d\sigma/dt$ for the process $\scat$.

\subsection{Scalar and Tensor Interactions}

\begin{figure}[h!]
\centering
\includegraphics[width=7cm]{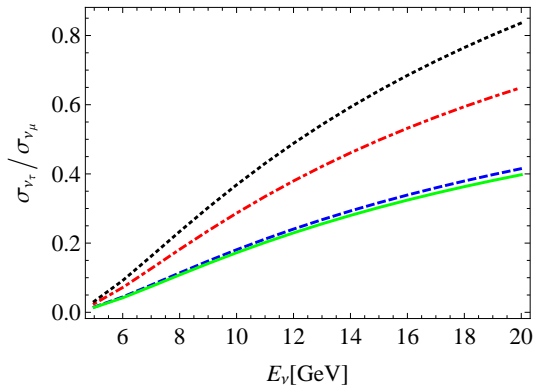}~~~
\caption{$S\pm T$ model: The ratio between the total cross section of $\scat$ to $\scatmu$ with Scalar-Tensor couplings. The green solid line corresponds to the standard model predictions $S_R=S_L=T_L=0$. The blue dashed, red dotdashed and black dotted lines correspond to $(S_R,S_L,T_L)=(-0.19, 0.68, 0.072),\;(1.98, 0.42, -0.13),\;(-1.87, -1.31, 0.18)$. }
\label{sigmaratio}
\end{figure}

\begin{figure}
\centering
\includegraphics[width=7cm]{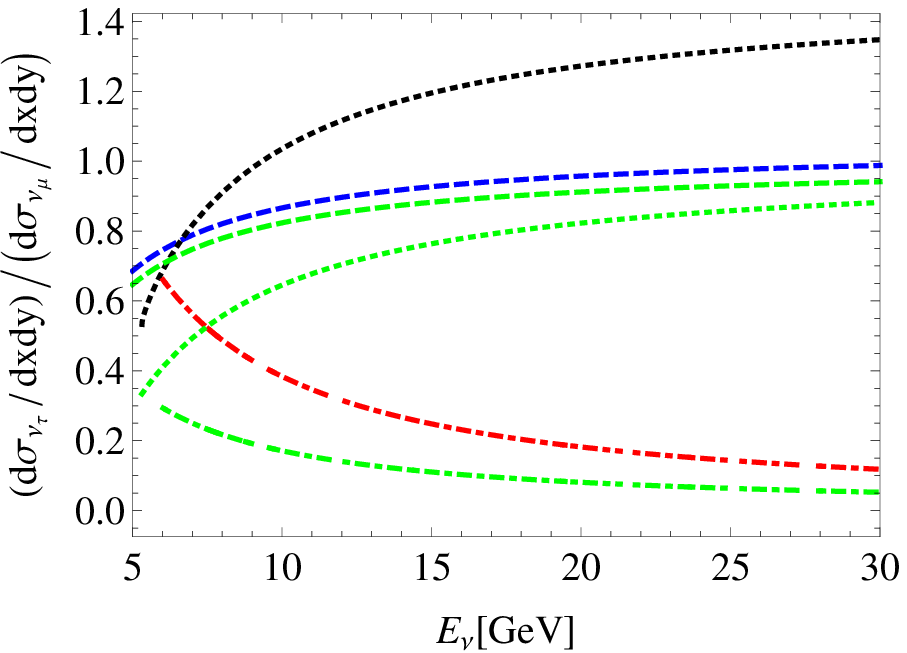}~~~
\caption{$S\pm T$ model: The ratio between the differential cross section $(d\sigma / dxdy)$ of $\scat$ to $\scatmu$ with  Scalar-Tensor couplings. The green lines correspond to the standard model predictions $S_R=S_L=T_L=0$. The blue, black, and red lines correspond to $(S_R,S_L,T_L)=(-0.19, 0.68, 0.072),\;(1.98, 0.42, -0.13),\;(-1.87, -1.31, 0.18)$. The blue and green dashed lines correspond to $(x,y)=(0.95,A+B)$. The black and green dotdashed lines correspond to $(x,y)=(0.475,(A+B)/2)$. The red and green dotted lines correspond to $(x,y)=(\frac{m_\tau^2}{2M(E_\nu -m_\tau)},A-B)$.}
\label{dsigmadxdyratio}
\end{figure}

\begin{figure}
\centering
\includegraphics[width=7cm]{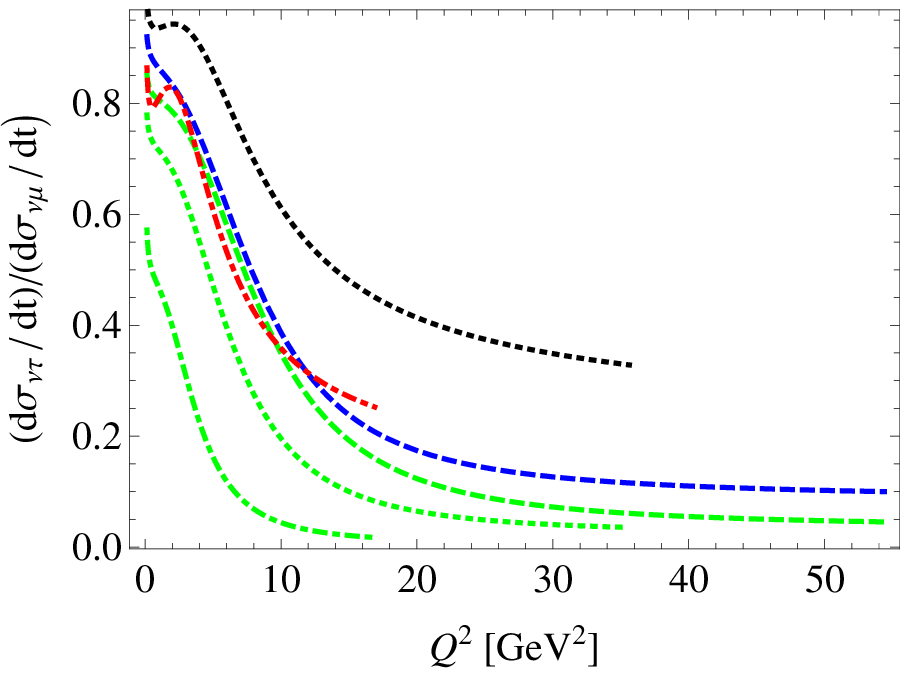}~~~
\caption{$S\pm T$ model:  The ratio between the differential cross section $(d\sigma / dt)$ of $\scat$ to $\scatmu$ in the Scalar-Tensor model. The green dashed, dotted and dotdashed lines correspond to the standard model predictions $S_R=S_L=T_L=0$ at $E_\nu = 30, 20, 10$ GeV, respectively. The blue dashed, black dotted, and red dotdashed lines correspond to $(S_R,S_L,T_L)=(-0.19, 0.68, 0.072),\;(1.98, 0.42, -0.13),\;(-1.87, -1.31, 0.18)$ at $E_\nu = 30, 20, 10$ GeV, respectively. The physical regions of the momentum transfer is taken to be $Q^2_-(W_{cut})\leq Q^2 \leq Q^2_+(W_{cut})$.}
\label{dsigmadtratio}
\end{figure}

\begin{figure}
\centering
\includegraphics[width=7cm]{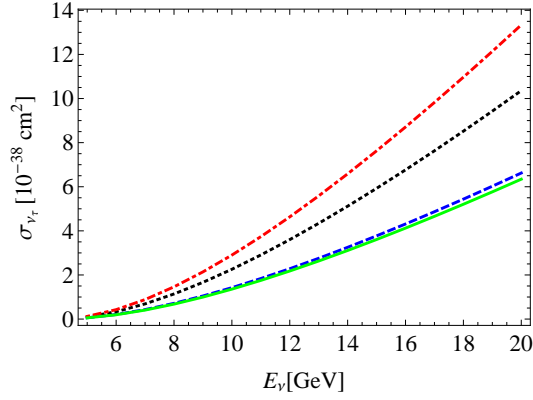}~~~
\caption{$S\pm T$ model: The total cross section of $\scat$ in the Scalar-Tensor model. The green solid line corresponds to the standard model predictions $S_R=S_L=T_L=0$. The blue dashed, black dotted and red dotdashed lines correspond to $(S_R,S_L,T_L)=(-0.19, 0.68, 0.072),\;(1.98, 0.42, -0.13),\;(-1.87, -1.31, 0.18)$. }
\label{sigmaratiotau}
\end{figure}

\begin{figure}
\centering
\includegraphics[width=7cm]{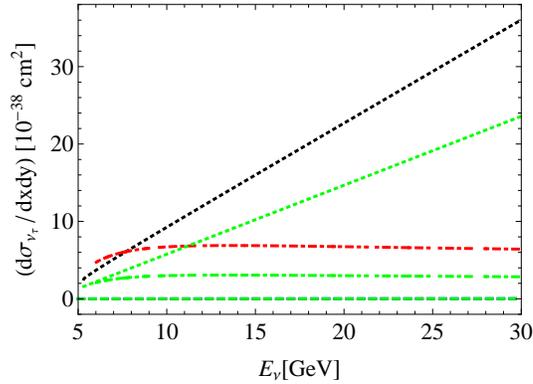}~~~
\caption{$S\pm T$ model: The differential cross section $(d\sigma / dxdy)$ of $\scat$ in the Scalar-Tensor model. The green lines correspond to the standard model predictions  $S_R=S_L=T_L=0$. The blue, black, and red lines correspond to $(S_R,S_L,T_L)=(-0.19, 0.68, 0.072),\;(1.98, 0.42, -0.13),\;(-1.87, -1.31, 0.18)$. The blue and green dashed lines correspond to $(x,y)=(0.95,A+B)$. The black and green dotdashed lines correspond to $(x,y)=(0.475,(A+B)/2)$. The red and green dotted lines correspond to $(x,y)=(\frac{m_\tau^2}{2M(E_\nu -m_\tau)},A-B)$.}
\label{dsigmadxdyratiotau}
\end{figure}

\begin{figure}
\centering
\includegraphics[width=5cm]{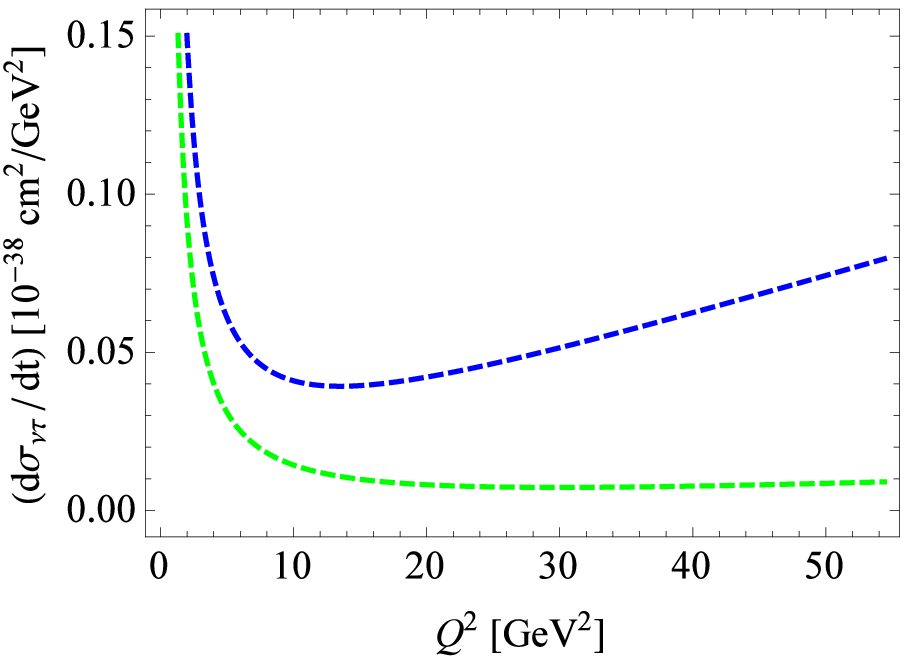}~~~
\includegraphics[width=5cm]{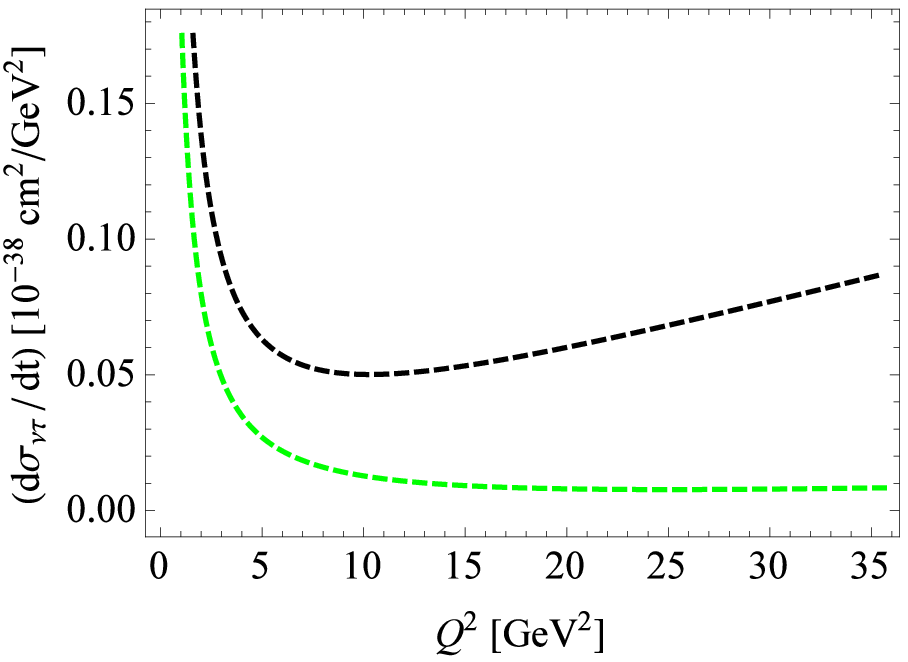}~~~
\includegraphics[width=5cm]{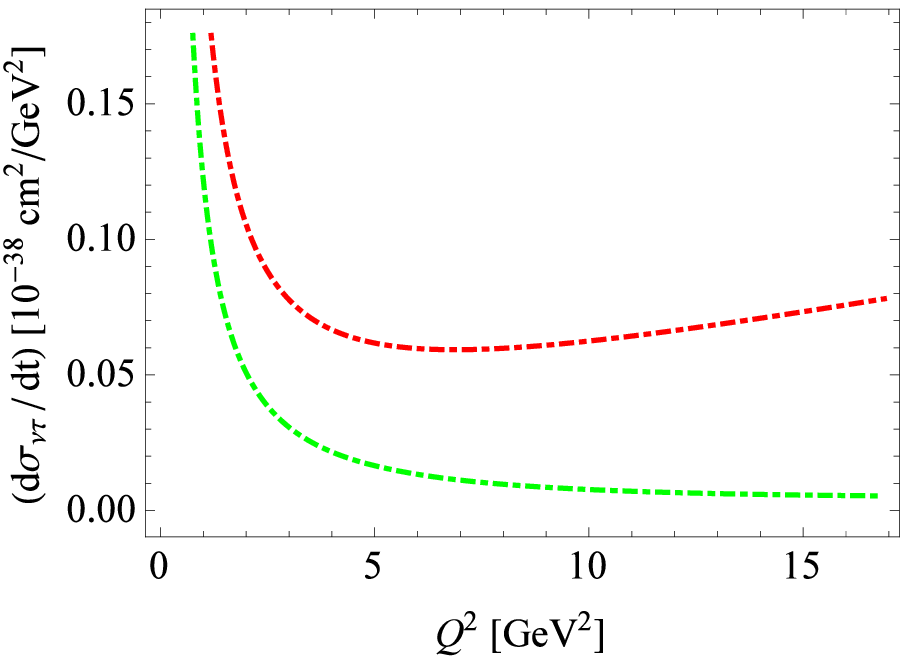}~~~
\caption{$S\pm T$ model:  The differential cross section $(d\sigma / dt)$ of $\scat$ in the Scalar-Tensor model. The green dashed, dotted and dotdashed lines correspond to the standard model predictions $S_R=S_L=T_L=0$ at $E_\nu = 30 $(left), $20$(middle) and $10 $(right) GeV, respectively. The blue dashed, black dotted, and red dotdashed lines correspond to $(S_R,S_L,T_L)=(-1.87, -1.31, 0.18)$ at $E_\nu = 30 $(left), $20$(middle) and $10 $(right) GeV, respectively. The physical regions of the momentum transfer is taken to be $Q^2_-(W_{cut})\leq Q^2 \leq Q^2_+(W_{cut})$.}
\label{dsigmadtratiotau}
\end{figure}

The ratio of the total cross section is shown in Fig.~\ref{sigmaratio} while the ratio of the differential cross sections $d \sigma/dxdy$ and $d\sigma/dt$ are given in Figs.~(\ref{dsigmadxdyratio}, \ref{dsigmadtratio}). The impact of the new physics is clearly detectable in the ratio of the total cross section and the differential cross sections. The new physics effect is also observable in the total cross section, $d \sigma/dxdy$, and $d\sigma/dt$ for the process $\scat$, as shown in Figs.~(\ref{sigmaratiotau}, \ref{dsigmadxdyratiotau}, \ref{dsigmadtratiotau}).

\subsection{Explicit Leptoquark Model}

\begin{figure}
\centering
\includegraphics[width=7cm]{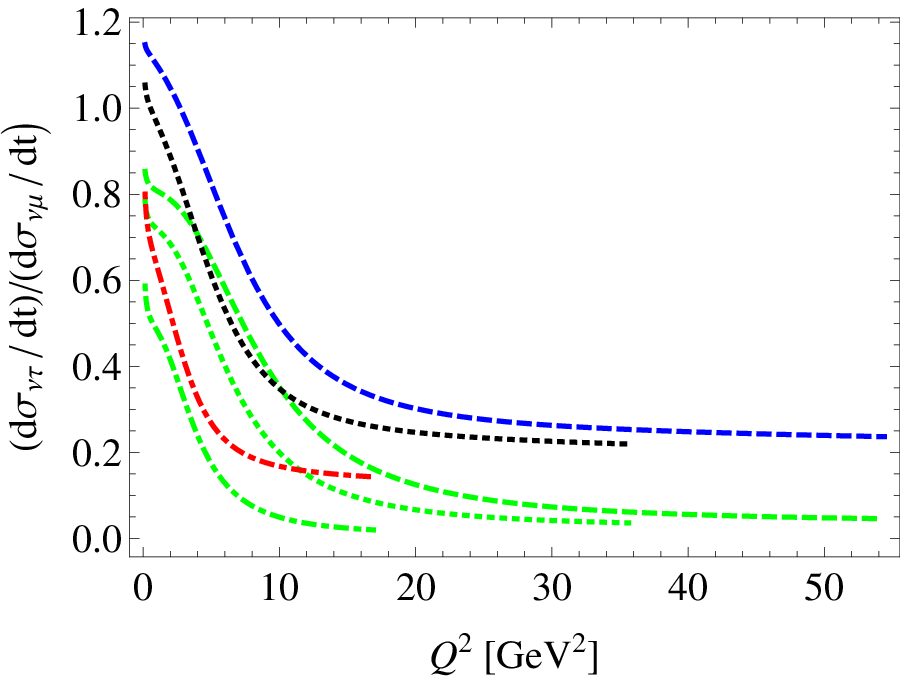}~~~
\caption{Leptoquark:  The ratio between the differential cross section $(d\sigma / dt)$ of $\scat$ to $\scatmu$ in the Leptoquark model with $S_L (m_{\rm LQ})=\pm 4 \; T_L (m_{\rm LQ})$ at $m_{\rm LQ}=1000$ GeV. The green dashed, dotted and dotdashed lines correspond to the standard model predictions $S_R=S_L=T_L=0$ at $E_\nu = 30, 20, 10$ GeV, respectively. The blue, black, and red lines correspond to $(Re[S_L(m_{LQ})],Im[S_L(m_{LQ})], Re[T_L(m_{LQ})], Im[T_L(m_{LQ})])=(0.56, 0.60, 0.14, 0.15)$ at $E_\nu = 30, 20, 10$ GeV, respectively. The physical regions of the momentum transfer is taken to be $Q^2_-(W_{cut})\leq Q^2 \leq Q^2_+(W_{cut})$.}
\label{dsigmadtratio2}
\end{figure}

\begin{figure}
\centering
\includegraphics[width=5cm]{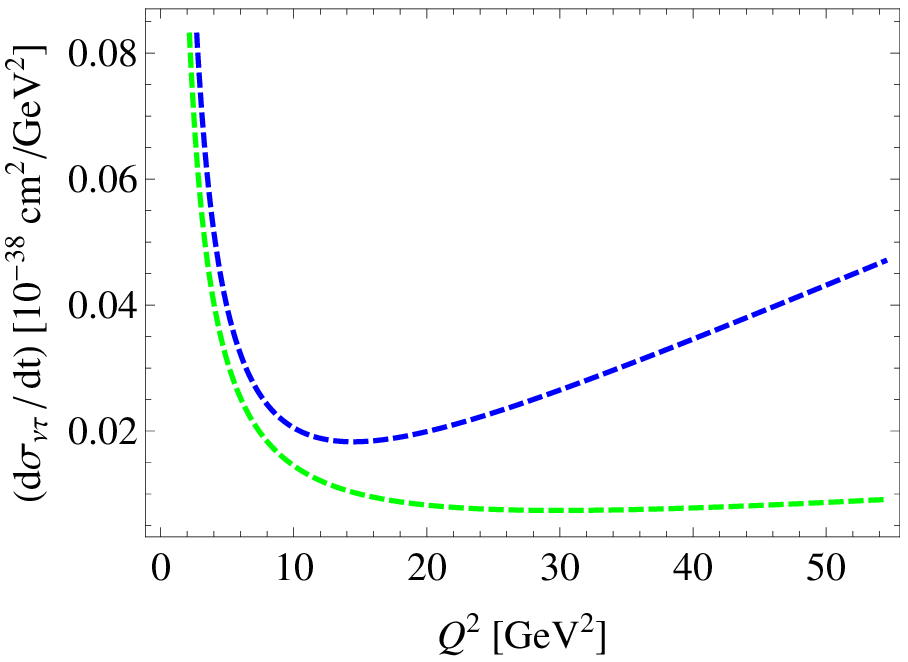}~~~
\includegraphics[width=5cm]{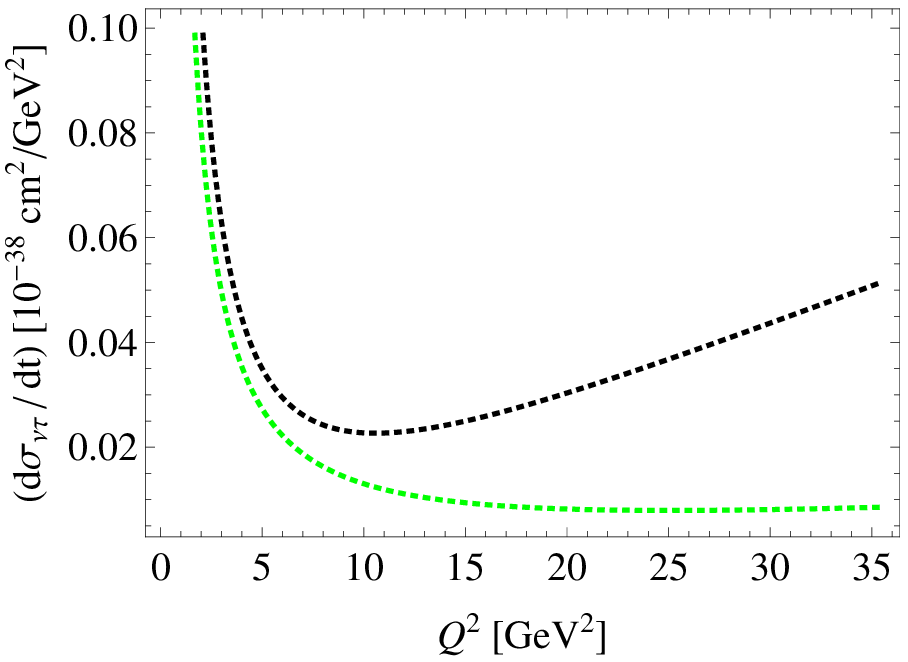}~~~
\includegraphics[width=5cm]{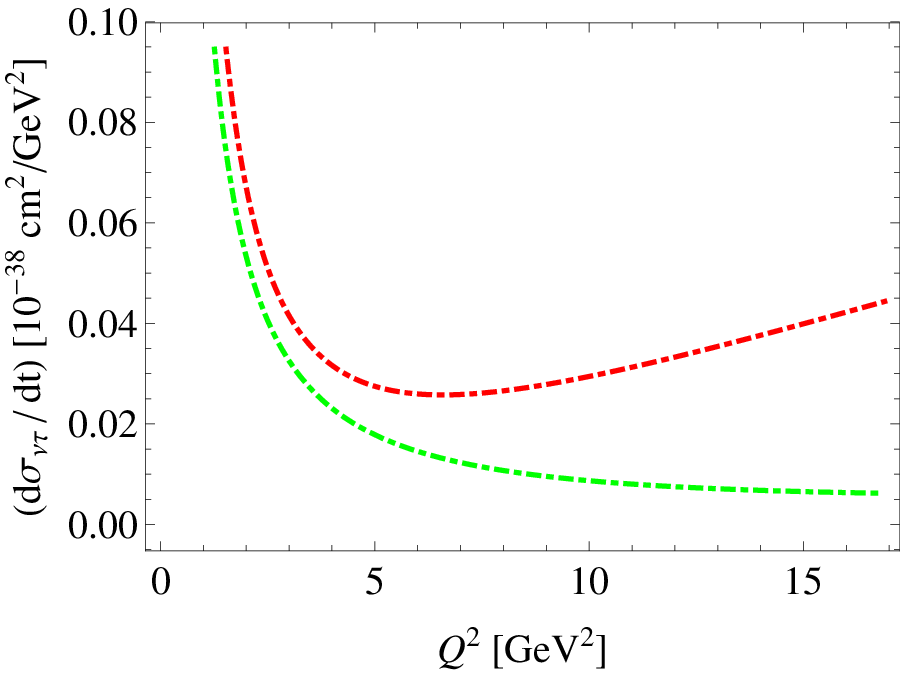}~~~
\caption{Leptoquark:  The differential cross section $(d\sigma / dt)$ of $\scat$ in the Leptoquark model with $S_L (m_{\rm LQ})=\pm 4 \; T_L (m_{\rm LQ})$ at $m_{\rm LQ}=1000$ GeV. The green dashed, dotted and dotdashed lines correspond to the standard model predictions $S_R=S_L=T_L=0$ at $E_\nu = $ 30(left), 20(middle) and 10(right) GeV, respectively. The blue, black, and red lines correspond to $(Re[S_L(m_{LQ})],Im[S_L(m_{LQ})], Re[T_L(m_{LQ})], Im[T_L(m_{LQ})])=(0.56, 0.60, 0.14, 0.15)$ at $E_\nu = $ 30(left), 20(middle) and 10(right) GeV, respectively. The physical regions of the momentum transfer is taken to be $Q^2_-(W_{cut})\leq Q^2 \leq Q^2_+(W_{cut})$.}
\label{dsigmadtratio2tau}
\end{figure}

\begin{figure}
\centering
\includegraphics[width=7cm]{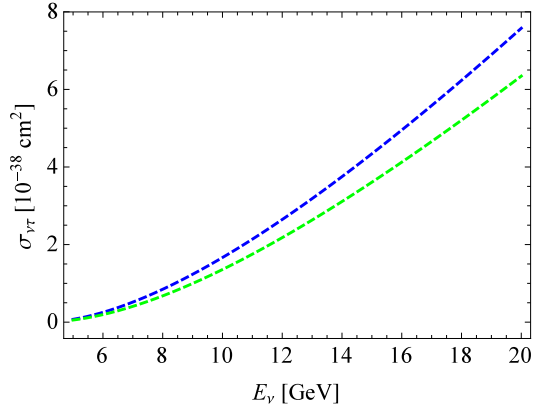}~~~
\caption{Leptoquark:  The total cross section in the Leptoquark model with $S_L (m_{\rm LQ})=\pm 4 \; T_L (m_{\rm LQ})$ at $m_{\rm LQ}=1000$ GeV. The green line corresponds to the standard model predictions $S_R=S_L=T_L=0$. The blue line corresponds to $(Re[S_L(m_{LQ})],Im[S_L(m_{LQ})], Re[T_L(m_{LQ})], Im[T_L(m_{LQ})])=(0.56, 0.60, 0.14, 0.15)$. The physical regions of the momentum transfer is taken to be $Q^2_-(W_{cut})\leq Q^2 \leq Q^2_+(W_{cut})$.}
\label{dsigmadtratio2tau2}
\end{figure}

Here we take $m_{\rm LQ}=1$ TeV.
In Figs.~(\ref{dsigmadtratio2}, \ref{dsigmadtratio2tau}, \ref{dsigmadtratio2tau2}), we show the differential cross section $d\sigma/dt$, its ratio, and the total cross  section for the particular models $S_L (m_{\rm LQ})=\pm 4 \; T_L (m_{\rm LQ})$. The impact of the new physics is clearly detectable.

\subsection{$V\pm A$ Interactions}

\begin{figure}
\centering
\includegraphics[width=7cm]{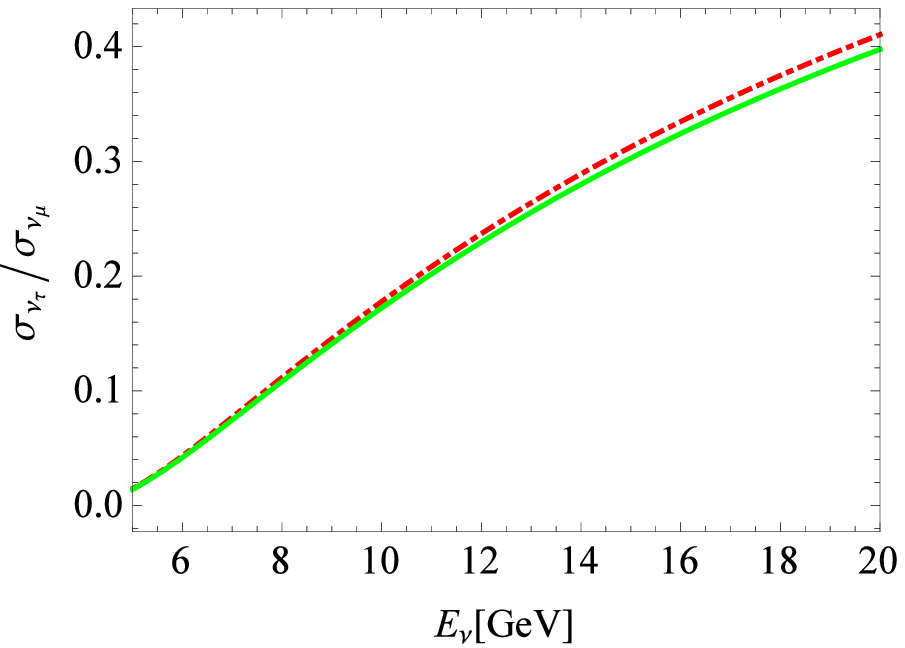}~~~
\caption{$V \pm A$ model: The ratio between the total cross section of $\scat$ to $\scatmu$ in the $V \pm A$ model. The green line corresponds to the standard model predictions $V_L = V_R =0$. The red dotdashed line corresponds to $(V_L , V_R)=(0.016, 0.006)$.}
\label{sigmaratioWp}
\end{figure}

\begin{figure}
\centering
\includegraphics[width=7cm]{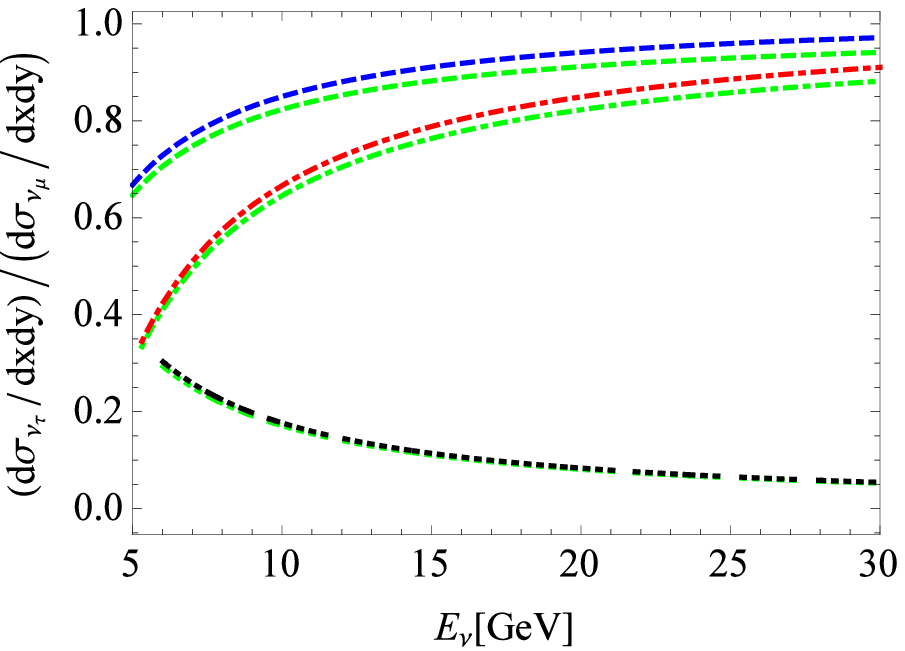}~~~
\caption{$V \pm A$ model: The ratio between the differential cross section $(d\sigma / dxdy)$ of $\scat$ to $\scatmu$ in the $V\pm A$ model. The green lines correspond to the standard model predictions $V_L = V_R =0$. The blue, black, and red lines correspond to $(V_L , V_R)=(0.016, 0.006)$. The blue and green dashed lines correspond to $(x,y)=(0.95,A+B)$. The red and green dotdashed lines correspond to $(x,y)=(0.475,(A+B)/2)$. The black and green dotted lines correspond $(x,y)=(\frac{m_\tau^2}{2M(E_\nu -m_\tau)},A-B)$. }
\label{dsigmadxdyratioWp}
\end{figure}

\begin{figure}
\centering
\includegraphics[width=7cm]{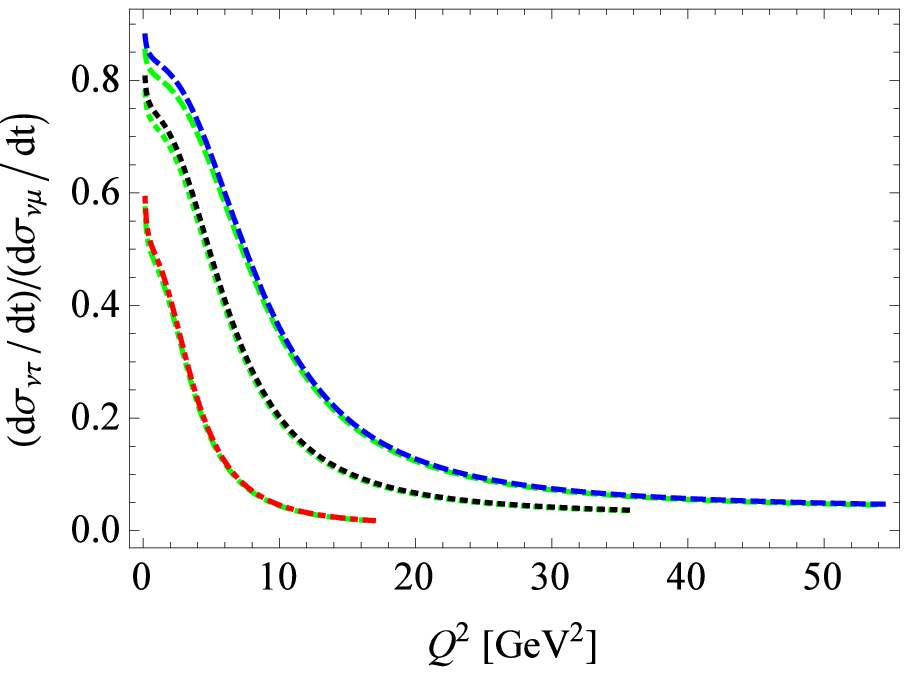}~~~
\caption{$V \pm A$ model:  The ratio between the differential cross section $(d\sigma / dt)$ of $\scat$ to $\scatmu$ in the $V\pm A$ model. The green lines correspond to the standard model predictions $V_L = V_R =0$. The blue, black, and red lines correspond to $(V_L , V_R)=(0.016, 0.006)$ at $E_\nu = 30, 20, 10$ GeV, respectively. The physical regions of the momentum transfer is taken to be $Q^2_-(W_{cut})\leq Q^2 \leq Q^2_+(W_{cut})$. }
\label{dsigmadtratioWp}
\end{figure}

\begin{figure}
\centering
\includegraphics[width=7cm]{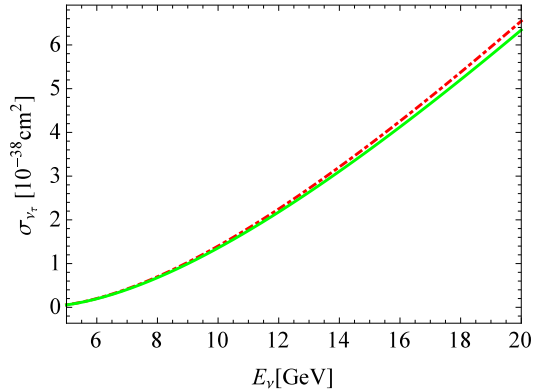}~~~
\caption{$V \pm A$ model:  The total cross section of $\scat$ in the $V \pm A$ model. The green line corresponds to the standard model predictions $V_L = V_R =0$. The red dotdashed line corresponds to $(V_L , V_R)=(0.016, 0.006)$.}
\label{sigmaratioWptau}
\end{figure}

\begin{figure}
\centering
\includegraphics[width=7cm]{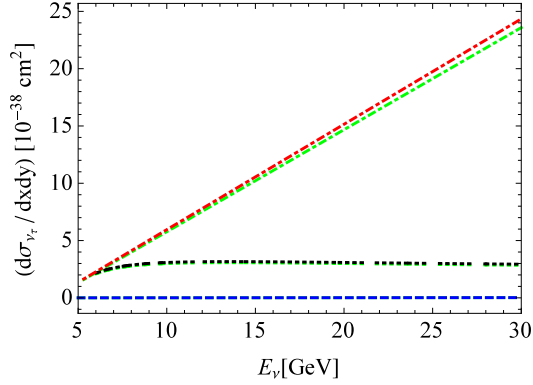}~~~
\caption{$V \pm A$ model:  The differential cross section $(d\sigma / dxdy)$ of $\scat$ in the $V\pm A$ model. The green lines correspond to the standard model predictions $V_L = V_R =0$. The blue, black, and red lines correspond to $(V_L , V_R)=(0.016, 0.006)$. The blue and green dashed lines correspond to $(x,y)=(0.95,A+B)$. The red and green dotdashed lines correspond to $(x,y)=(0.475,(A+B)/2)$. The black and green dotted lines correspond to $(x,y)=(\frac{m_\tau^2}{2M(E_\nu -m_\tau)},A-B)$. }
\label{dsigmadxdyratioWptau}
\end{figure}

\begin{figure}
\centering
\includegraphics[width=7cm]{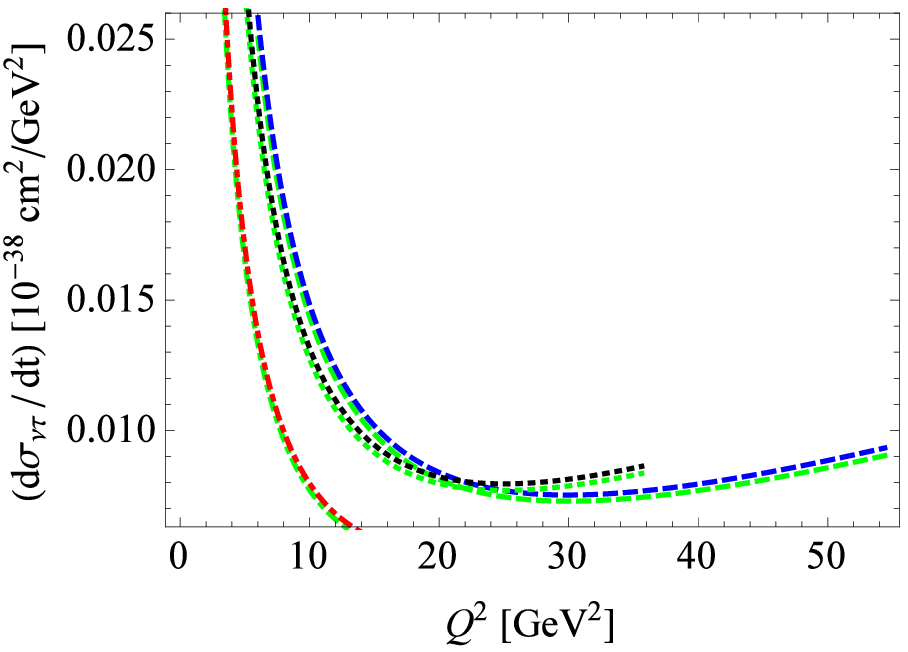}~~~
\caption{$V \pm A$ model:  The differential cross section $(d\sigma / dt)$ of $\scat$ in the $V\pm A$ model. The green lines correspond to the standard model predictions $V_L = V_R =0$. The blue, black, and red lines correspond to $(V_L , V_R)=(0.016, 0.006)$ at $E_\nu = 30, 20, 10$ GeV, respectively. The physical regions of the momentum transfer is taken to be $Q^2_-(W_{cut})\leq Q^2 \leq Q^2_+(W_{cut})$ }
\label{dsigmadtratioWptau}
\end{figure}

The ratio of the total cross section, $d\sigma/dxdy$, $d\sigma/dt$ are shown in Figs.~(\ref{sigmaratioWp}, \ref{dsigmadxdyratioWp}, \ref{dsigmadtratioWp}), respectively. The figures show that the effect of $V \pm A$ new physics is small in the neutrino cross section.
The new physics effect is small in the total cross section, $d \sigma/dxdy$, and $d\sigma/dt$ for the process $\scat$, as shown in Figs.~(\ref{sigmaratioWptau}, \ref{dsigmadxdyratioWptau}, \ref{dsigmadtratioWptau}).


\pagebreak

\section{Conclusion}
In this paper we discussed tests of lepton non-universal interactions through $\nu_{\tau}$ scattering. We adopted an effective Lagrangian description of new physics and considered  explicit leptoquark models for our calculations.
The parameters of the new physics were constrained by single pion and two pion $\tau$ decays,  $\tau^-(k_1) \rightarrow \nu_\tau (k_2)+\pi^-(q)$ and $\tau(p) \to \pi^-(p_1)+ \pi^0(p_2)+\nu_{\tau}(p_3)$ , which are well measured. 
We then discussed the ratio of the total and differential cross sections for the two deep inelastic scattering processes $\scat$ and $\scatmu$ as a probe of the new physics in the neutrino cross-section experiments. In the ratio of cross sections, the uncertainty of the parton distribution functions is expected to cancel out leading to precise results. In the effective Lagrangian framework we looked at models with scalar and tensor interactions. As an explicit realization of such models we considered leptoquark models where scalar and tensor couplings arise with relations between the couplings. 
Our results showed significant new physics effects, both in the total cross sections as well as in the differential distributions for
$\scat$, are allowed with the present constraints. These new physics effects could be observed at future proposed $\nu_{\tau}$ scattering experiments. We also considered vector-axial vector new physics operators in our analysis. The
results showed that the new physics effect is small in this case.

\section*{Acknowledgements}

This work was financially supported in part by the National Science Foundation under Grant No.NSF PHY-1414345 (A.D and H.L).

\section*{Appendix (A)}

Here, we give details of the calculations of the process $\dec$. In the rest frame of $\pi^-$ and $\pi^0$,
\bea
p=
\begin{pmatrix}
E, & \vec{P}\\
\end{pmatrix}
,
p_1=
\begin{pmatrix}
E_1, & \vec{P_1}\\
\end{pmatrix}
,
p_2=
\begin{pmatrix}
E_2, & -\vec{P_{1}}\\
\end{pmatrix}
,
p_3=
\begin{pmatrix}
E_3, & \vec{P}\\
\end{pmatrix}
,\\
k=p_1-p_2=
\begin{pmatrix}
E_1-E_2, & 2\vec{P_{1}}\\
\end{pmatrix}
,
q=p_1+p_2=
\begin{pmatrix}
E_1+E_2, & 0\\
\end{pmatrix},
\eea
and we define two variables,
\bea
m_{12}^2 &=& (p_1+p_2)^2=q^2,\nonumber\\
m_{23}^2 &=& (p_2+p_3)^2.
\eea
Then,
\begin{equation}
d\Gamma=\frac{1}{(2\pi)^3}\frac{1}{32m_{\tau}^3}Xdm_{12}^2dm_{23}^2,
\end{equation}\\
with
\begin{equation}
X=\frac{1}{2}\sum_{spin}|M_{SM}+M_{T}|^2=\frac{1}{2}\sum_{spin}|M_{SM}|^2+\frac{1}{2}\sum_{spin}|M_{T}|^2,
\end{equation}
where $M_{SM}$ and $M_{T}$ are given in Eqs.~(\ref{MSM}, \ref{MLQT}), and the cross terms are zero.

By averaging the spin, we  get
\begin{equation}
X_{SM}=4G_F^2 V_{ud}^2 F^2(Q^2)[2(k\cdot p)(k\cdot p_3)-k^2(p\cdot p_3)],
\end{equation}
 and,
\begin{eqnarray}
 X_{T}&=&16G_F^2 V_{ud}^2 T_L^2 \frac{F^2(Q^2)}{q^2} [(k\cdot q)^2(-p\cdot p_3)+2(k\cdot q)((k\cdot p_3)(p\cdot q)\nonumber\\
&&+(k\cdot p)(p_3\cdot q))-2q^2(k\cdot p)(k\cdot p_3)+k^2(q^2(p\cdot p_3)-2(p\cdot q)(p_3\cdot q))].\nonumber\\
\end{eqnarray}
All these $X's$ can be expressed in terms of $m_{12}^2$ and $m_{23}^2$ because
\begin{equation}
E_1=\frac{m_{12}^2-m_2^2+m_1^2}{2m_{12}},
\end{equation}
\begin{equation}
E_2=\frac{m_{12}^2-m_1^2+m_2^2}{2m_{12}},
\end{equation}
\begin{equation}
E_3=\frac{M^2-m_{12}^2-m_3^2}{2m_{12}},
\end{equation}
\begin{equation}
E=\frac{M^2+m_{12}^2-m_3^2}{2m_{12}},
\end{equation}
\begin{equation}
2 \vec{p_1}\cdot \vec{p}=m_{23}^2-m_2^2-m_3^2-2E_2E_3,
\end{equation}
\begin{equation}
|\vec{p}|=\sqrt{E^2-M^2},
\end{equation}
\begin{equation}
|\vec{p_1}|=\sqrt{E_1^2-m_1^2},
\end{equation}
with $M=m_{\tau}=1.77$ GeV, $m_1=m_{\pi^-}=0.140$ GeV, $m_2=m_{\pi^0}=0.135$ GeV, $m_3=m_{\nu_{\tau}}=0$.

Let's work on the SM case first and set
 \begin{align}
 A_1 &=k\cdot p=E(E_1-E_2)-2\vec{p_1}\cdot\vec{p}\nonumber\\
     &=E(E_1-E_2)+2E_2E_3+m_2^2+m_3^2-m_{23}^2,\\
 A_2 &=k\cdot p_3=E_3(E_1-E_2)-2\vec{p_1}\cdot\vec{p}\nonumber\\
     &=E_3(E_1-E_2)+2E_2E_3+m_2^2+m_3^2-m_{23}^2,\\
 A_3 &=k^2=(E_1-E_2)^2-4\vec{p_1}^2\nonumber\\
     &=(E_1-E_2)^2-4(E_1^2-m_1^2),\\
 A_4 &=p\cdot p_3=E_1E_3-\vec{p}^2\nonumber\\
     &=E_1E_3-E^2+M^2.
 \end{align}
Then, 
\begin{equation}
X_{SM}=4G_F^2 V_{ud}^2 F^2(Q^2)[2A_1A_2-A_3A_4].
\end{equation}
Now, let us integrate $X_{SM}$ by $m_{23}^2$ within the limits
\begin{equation}
(m_{23}^2)_{max}=(E_2+E_3)^2-\left( \sqrt{E_2^2-m_2^2}-\sqrt{E_3^2-m_3^2}\right)^2,
\end{equation}
\begin{equation}
(m_{23}^2)_{min}=(E_2+E_3)^2-\left( \sqrt{E_2^2-m_2^2}+\sqrt{E_3^2-m_3^2}\right)^2,
\end{equation} 
where $m_1=m_2=m_{\pi}$, $m^2_{12}=Q^2$ and $M=m_{\tau}$. One gets
\begin{equation}
\Gamma_{SM}=\frac{4G^2_Fm^5_{\tau}}{96(2\pi)^3}\frac{cos\theta_c}{2}\int_{Q^2_{min}}^{Q^2_{max}}\frac{dQ^2}{m^2_{\tau}}F^2(Q^2)\left( 1-\frac{Q^2}{m^2_{\tau}}\right)^2 \left( 1+2\frac{Q^2}{m^2_{\tau}}\right) \left( 1-\frac{4m^2_{\pi}}{Q^2}\right)^{3/2}.
\end{equation}
Now, we can integrate over $m_{12}^2$ within the limits
\begin{eqnarray}
Q^2_{max}&=&(m_{12}^2)_{max}=(M-m_1)^2,\nonumber\\
Q^2_{min}&=&(m_{12}^2)_{min}=(m_1+m_2)^2. 
\end{eqnarray}
Now let's work on the tensor leptoquark case, we set 
 \begin{align}
 B_1 &=k\cdot q=E^2_1-E^2_2,\\
 B_2 &=p\cdot p_3=EE_3-\vec{p}^2,\\
 B_3 &=k\cdot p_3=E_3(E_1-E_2)-2\vec{p}\cdot\vec{p_1},\\
 B_4 &=p\cdot q=E(E_1+E^2),\\
 B_5 &=k\cdot p_3=E(E_1-E_2)-2\vec{p}\cdot\vec{p_1},\\
 B_6 &=P_3\cdot q=E_3(E_1+E_2),\\
 B_7 &=q^2=(E_1+E_2)^2,\\
 B_8 &=k^2=(E_1-E_2)^2-4\vec{p_1}^2 .
\end{align}
Then,
\begin{equation}
X_{T}=8 G_F^2 V_{ud}^2 T_L^2 F_T^2 (q^2) \left[-B_1^2 B_3 + 2 B_1 (B_3 B_4 + B_1 B_6)-2 B_7 B_1 B_5 + B_8 (B_7 B_2 - 2 B_4 B_6) \right] .
\end{equation}
Numerically, we can get $\Gamma_{T}=3.43\times 10^{-12}T_L^2$ GeV.

\section*{Appendix (B)}

In the decay process $\tau^- \rightarrow \nu_\tau +\pi^-$, the SM hadronic current is given in Eq.~\ref{SM-current}. By multiplying the current by $q^\mu=p^\mu_d + p^\mu_u$, one can find the NP scalar current given by
\beq
\langle 0|\bar{d}(A_S-B_S \gamma^5)u|\pi(q)\rangle = \frac{i\sqrt{2}f_\pi m_\pi^2}{m_u+m_d}B_S.
\eeq
If one multiplies the SM current by $k^\mu=p^\mu_d - p^\mu_u$, the scalar current will be 
\beq
\langle 0|\bar{d}(A_S-B_S \gamma^5)u|\pi(q)\rangle = i\sqrt{2}f_\pi (m_u+m_d)B_S.
\eeq
Now, by multiplying the two equations above and taking the square root, we end up with scalar current that is independent of the quark masses
\beq
\langle 0|\bar{d}(A_S-B_S \gamma^5)u|\pi(q)\rangle = i\sqrt{2}f_\pi m_\pi B_S.
\eeq

In the process $\tau^- (p) \to \pi^-(p_1)+ \pi^0(p_2)+\nu_{\tau}(p_3)$, the NP tensor current is given in Eq.~\ref{current2}. Here $p_1^\mu = p^\mu_d + p^\mu_q$ and $p_2^\mu = p^\mu_u - p^\mu_q$, where $p_u$ and $p_d$ are the momenta of the up and down quarks that come from the tau decay, and $(p_q, -p_q)$ are the momenta of the quark-antiquark pair from the vacuum that pair up with the  up and down quarks to form $\pi^{0}$ and $\pi^-$. By multiplying the current by $q^\mu=p^\mu_1 + p^\mu_2=p^\mu_d - p^\mu_u$ and using the equation of motion, in the isospin symmetry limit, one gets the form factor
\begin{equation}
F_T=-i\frac{(m_u+m_d)}{q^2}F.
\end{equation}
If one multiplies the tensor current by $k^\mu=p^\mu_1 - p^\mu_2=p^\mu_d - p^\mu_u+2p^\mu_q$, the form factor will be given by
\beq
F_T=\frac{-i F}{(m_d + m_u)-\left( 1-2\frac{p_q \cdot k}{k^2}\right) \left(\frac{m_{\pi^-}^2-m_{\pi^0}^2}{m_d-m_u} \right) }.
\eeq
Now if the $\pi \pi$ is dominantly coming from a vector resonance then we can expect that the distribution of the momenta of the quarks inside the resonance will be peaked around
$p_u=p_d$. In this limit the second term in the denominator above vanishes as $\left( 1-2\frac{p_q \cdot k}{k^2}\right)=0$ .
Hence,by taking the second term in the denominator small, we get 
\begin{equation}
F_T=-i\frac{1}{(m_u+m_d)}F.
\end{equation}
Now, by multiplying the two equations above and taking the square root,  the form factor will be independent of the quark masses
\begin{equation}
F_T=\frac{-i F}{\sqrt{q^2}}.
\end{equation}



\begin{thebibliography}{99}


\bibitem{RDexpt1}   J.~P.~Lees {\it et al.}  [BaBar Collaboration],
  Phys.\ Rev.\ Lett.\  {\bf 109}, 101802 (2012)
  [arXiv:1205.5442 [hep-ex]].

\bibitem{RDexpt2} J.~P.~Lees {\it et al.}  [BaBar Collaboration],
  Phys.\ Rev.\ D {\bf 88}, 072012 (2013)
  [arXiv:1303.0571 [hep-ex]].

\bibitem{RDtheory} S.~Fajfer, J.~F.~Kamenik and I.~Nisandzic,
  [arXiv:1203.2654 [hep-ph]];
Y.~Sakaki and H.~Tanaka,
  [arXiv:1205.4908 [hep-ph]].

\bibitem{dattaD} See for instance,
A.~Datta, M.~Duraisamy and D.~Ghosh,
  Phys.\ Rev.\ D {\bf 86}, 034027 (2012)
  [arXiv:1206.3760 [hep-ph]];
M.~Duraisamy and A.~Datta,
  JHEP {\bf 1309}, 059 (2013)
  [arXiv:1302.7031 [hep-ph]];
M.~Duraisamy, P.~Sharma and A.~Datta,
  Phys.\ Rev.\ D {\bf 90}, 074013 (2014)
  [arXiv:1405.3719 [hep-ph]];
  S.~Shivashankara, W.~Wu and A.~Datta,
  arXiv:1502.07230 [hep-ph].
  

\bibitem{Bhattacharya:2014wla} 
  B.~Bhattacharya, A.~Datta, D.~London and S.~Shivashankara,
  Phys.\ Lett.\ B {\bf 742}, 370 (2015)
  [arXiv:1412.7164 [hep-ph]].


\bibitem{RKexpt} R.~Aaij {\it et al.}  [LHCb Collaboration],
  Phys.\ Rev.\ Lett.\  {\bf 113}, 151601 (2014)
  [arXiv:1406.6482 [hep-ex]].

\bibitem{RKtheory} G.~Hiller and F.~Kruger,
  Phys.\ Rev.\ D {\bf 69}, 074020 (2004)
  [arXiv:hep-ph/0310219];
C.~Bobeth, G.~Hiller and G.~Piranishvili,
  JHEP {\bf 0712}, 040 (2007)
  [arXiv:0709.4174 [hep-ph]];
C.~Bouchard {\it et al.}  [HPQCD Collaboration],
  Phys.\ Rev.\ Lett.\  {\bf 111}, no. 16, 162002 (2013)
  [Erratum-ibid.\  {\bf 112}, no. 14, 149902 (2014)]
  [arXiv:1306.0434 [hep-ph]].



\bibitem{BKmumuhadunc}
See for example
A.~K.~Alok, A.~Dighe, D.~Ghosh, D.~London, J.~Matias, M.~Nagashima and A.~Szynkman,
  JHEP {\bf 1002}, 053 (2010)
  [arXiv:0912.1382 [hep-ph]];
A.~K.~Alok, A.~Datta, A.~Dighe, M.~Duraisamy, D.~Ghosh and D.~London,
  JHEP {\bf 1111}, 121 (2011)
  [arXiv:1008.2367 [hep-ph]],
  JHEP {\bf 1111}, 122 (2011)
  [arXiv:1103.5344 [hep-ph]];

\bibitem{Aaij:2013qta}
  R.~Aaij {\it et al.}  [LHCb Collaboration],
  Phys.\ Rev.\ Lett.\  {\bf 111} (2013) 191801
  [arXiv:1308.1707 [hep-ex]].

\bibitem{BKmumuNP}
See for example:
    A.~Datta, M.~Duraisamy and D.~Ghosh,
  Phys.\ Rev.\ D {\bf 89} (2014) 071501
  [arXiv:1310.1937 [hep-ph]];

\bibitem{Kodama:2007aa} 
  K.~Kodama {\it et al.}  [DONuT Collaboration],
  Phys.\ Rev.\ D {\bf 78}, 052002 (2008)
  [arXiv:0711.0728 [hep-ex]].


\bibitem{ourpapers}
A.~Rashed, P.~Sharma and A.~Datta,
  Nucl.\ Phys.\ B {\bf 877}, 662 (2013)
  [arXiv:1303.4332 [hep-ph]];
A.~Rashed, M.~Duraisamy and A.~Datta,
  Phys.\ Rev.\ D {\bf 87}, no. 1, 013002 (2013)
  [arXiv:1204.2023 [hep-ph]].

\bibitem{SHIP}
E.~Graverini, N.~Serra and B.~Storaci,
  arXiv:1503.08624 [hep-ex];
D.~Gorbunov, A.~Makarov and I.~Timiryasov,
  Phys.\ Rev.\ D {\bf 91}, no. 3, 035027 (2015)
  [arXiv:1411.4007 [hep-ph]].
  
  \bibitem{ccLag}
  T.~Bhattacharya, V.~Cirigliano, S.~D.~Cohen, A.~Filipuzzi, M.~Gonzalez-Alonso, M.~L.~Graesser, R.~Gupta and H.~-W.~Lin,
  Phys.\ Rev.\ D {\bf 85}, 054512 (2012)
  [arXiv:1110.6448 [hep-ph]];
  C.~-H.~Chen and C.~-Q.~Geng,
  Phys.\ Rev.\ D {\bf 71}, 077501 (2005)
  [hep-ph/0503123].
  
  
  
  \bibitem{dattaKK}
  A.~Datta, P.~J.~O'Donnell, Z.~H.~Lin, X.~Zhang and T.~Huang,
  Phys.\ Lett.\ B {\bf 483}, 203 (2000)
  [hep-ph/0001059].
  





\bibitem{kretzer}
S. Kretzer and M. H. Reno, Phys. Rev. D{\bf 66}(2002)113007.

\bibitem{albright}
C. H. Albright and C. Jarlskog, Nucl. Phys. B{\bf 84}(1975)467.






\bibitem{Buchmuller:1986zs} 
  W.~Buchmuller, R.~Ruckl and D.~Wyler,
  Phys.\ Lett.\ B {\bf 191}, 442 (1987)
  [Erratum-ibid.\ B {\bf 448}, 320 (1999)].
  
  
\bibitem{Queiroz:2014pra} 
  F.~S.~Queiroz, K.~Sinha and A.~Strumia,
  Phys.\ Rev.\ D {\bf 91}, no. 3, 035006 (2015)
  [arXiv:1409.6301 [hep-ph]].
  
  
  \bibitem{laglq2}
  W. Buchmuller, R. Ruckl and D. Wyler, Phys.Lett. B 191 (1987) 442.
  
  
  
  
  


  

%
%
%

  
  \bibitem{Dorsner:2013tla} 
  I.~Dorsner, S.~Fajfer, N.~Kosnik and I.~Nisandzic,
  JHEP {\bf 1311}, 084 (2013)
  [arXiv:1306.6493 [hep-ph]].
  











\bibitem{Hagiwara:2003di} 
  K.~Hagiwara, K.~Mawatari and H.~Yokoya,
  Nucl.\ Phys.\ B {\bf 668}, 364 (2003)
  [Erratum-ibid.\ B {\bf 701}, 405 (2004)]
  [hep-ph/0305324].










\bibitem{Chetyrkin:1997dh} 
  K.~G.~Chetyrkin,
  Phys.\ Lett.\ B {\bf 404}, 161 (1997)
  [hep-ph/9703278].



\bibitem{Gracey:2000am} 
  J.~A.~Gracey,
  Phys.\ Lett.\ B {\bf 488}, 175 (2000)
  [hep-ph/0007171].







\bibitem{pdg}
K. Nakamura et al. (Particle Data Group), J. Phys. G 37, 075021 (2010)
and 2011 partial update for the 2012 edition.


\bibitem{rad} 
  M.~Davier, A.~Hocker and Z.~Zhang,
  Rev.\ Mod.\ Phys.\  {\bf 78}, 1043 (2006)
  [hep-ph/0507078].

\bibitem{barish} 
  B.~C.~Barish,
  In *Stanford 1989, Proceedings, Study of tau, charm and J/psi physics* 113-126 and Caltech Pasadena - CALT-68-1580 (89,rec.Oct.) 14 p


%
%
%
%
%



\bibitem{tau-decay}
 J. H. Kuhn, A. Santamaria, Zeitschrift fur Physik C - Particles and Fields 1990, Volume 48, Issue 3, pp 445-452 

\bibitem{Bernicha:1995rh} 
  A.~Bernicha, G.~Lopez Castro and J.~Pestieau,
  Phys.\ Rev.\ D {\bf 53}, 4089 (1996)
  [hep-ph/9510435].



\bibitem{pdg2pi}
K.A. Olive et al. (Particle Data Group), Chin. Phys. C, 38, 090001 (2014). 


%
%







%
%
%
  







%
%











\end{thebibliography}
\end{document}